\documentclass[oldversion]{aa} 
\usepackage{graphicx}
\usepackage{longtable}
\usepackage{txfonts}
\usepackage{rotating}
\usepackage{lscape}
\newcommand{\gsim}{\;\lower.6ex\hbox{$\sim$}\kern-7.75pt\raise.65ex\hbox{$>$}\;}
\newcommand{\lsim}{\;\lower.6ex\hbox{$\sim$}\kern-7.75pt\raise.65ex\hbox{$<$}\;}

\begin{document}
\title{The Na-O anticorrelation in horizontal branch stars. II. NGC~1851
\thanks{Based on observations collected at 
ESO telescopes under program 386.D-0086}
}

\author{
R.G. Gratton\inst{1},
S. Lucatello\inst{1},
E. Carretta\inst{2},
A. Bragaglia\inst{2},
V. D'Orazi\inst{1},
Y. Al Momany\inst{1,3},
A. Sollima\inst{1},
M. Salaris\inst{4}
\and
S. Cassisi\inst{5}}

\authorrunning{R.G. Gratton}
\titlerunning{Na-O in HB stars of NGC~1851}

\offprints{R.G. Gratton, raffaele.gratton@oapd.inaf.it}

\institute{
INAF-Osservatorio Astronomico di Padova, Vicolo dell'Osservatorio 5, I-35122
 Padova, Italy
\and
INAF-Osservatorio Astronomico di Bologna, Via Ranzani 1, I-40127, Bologna, Italy
\and
European Southern Observatory, Alonso de Cordova 3107, Vitacura, Santiago, Chile 
\and
Astrophysics Research Institute, Liverpool John Moores University, United Kingdom
\and
INAF-Osservatorio Astronomico di Teramo, Via Collurania, Teramo, Italy}

\date{}
\abstract{The chemical composition of horizontal branch (HB) stars might help to 
clarify the formation history of individual globular clusters (GCs). We studied the 
Na-O anti-correlation from moderately high resolution spectra for 91
stars on the bimodal HB of NGC~1851; in addition we observed 13 stars on the lower
red giant branch (RGB). In our HB sample, 35 stars are on the blue HB
(BHB), one is an RR Lyrae, and 55 stars are on the red HB (RHB).
The ratio of BHB to RHB stars is close to the total in the cluster (35 and 54\%,
respectively), while RR Lyrae variables are under-represented, (they are $\sim 12$\% 
of the NGC~1851 stars). We also derived abundances for He and N in BHB stars. For 
RHB stars we derived Ba abundances and a few interesting upper limits for N. 
The RHB stars clearly separate into two groups: the vast majority are O-rich and
Na-poor, while about 10-15\% are Na-rich and moderately O-poor. Most (but not all)
Na-rich RHB stars are also Ba-rich and there is an overall correlation between Na 
and Ba abundances within the RHB. The group of Ba-rich RHB stars resides on the 
warmer edge and includes $\sim 10$\% of the RHB stars. We propose that they are 
the descendant of the stars on the RGB sequence with very red $v-y$\ colour. This 
sequence is known also to consist of Ba and perhaps CNO-rich stars and consistently 
includes $\sim 5-10$\% of the RGB stars of NGC1851. However, the upper 
limit we obtain for N ([N/Fe]$<1.55$) for one of the Ba-rich stars coupled with
the low C-abundances for RGB Ba-rich stars from the literature suggests that 
the total CNO might not be particularly high ([(C+N+O)/Fe]$\leq 0.15$). The other Na-rich RHB stars are also
at the warm edge of the RHB and the only RR Lyrae is Na-rich and moderately O-poor. We also find a Na-O 
anticorrelation among BHB stars, partially overlapping that found among RHB stars, 
though generally BHB stars are more Na-rich and O-poor. However, there is no clear 
correlation between temperature and Na and O abundances within the BHB. The average He 
abundance in BHB stars is Y=$0.29\pm 0.05$, which excludes a large population of
extremely He-rich stars from our sample. N abundances are quite uniform at 
[N/Fe]=$1.16\pm 0.14$\ among BHB stars, with a 
small trend with temperature. This value is consistent with normal CNO abundance
and excludes that BHB stars are very CNO-rich: this leaves an age spread of 
$\sim 1.5$~Gyr as the only viable explanation for the split of the SGB. To help 
clarifying the formation history of NGC~1851, we computed synthetic HB's trying to 
identify which HB stars are the descendant of the bright and faint subgiant branch 
(b-SGB and f-SGB) stars identified by Milone et al. (2008), with respectively 2/3 and 1/3 
of the stars of NGC~1851. 
While most BHB stars likely descend from f-SGB stars and are older, and most RHB 
stars from b-SGB ones and are younger, the correspondence is probably not one-to-one. 
In particular, the Ba-rich RHB stars should be less massive than the remaining RHB 
stars, and the location of their progenitors on the SGB is uncertain. If they descend from
f-SGB stars, number counts 
then require that RR Lyrae variables and possibly some mild BHB stars descend from 
b-SGB stars; this suggestion is supported by a few circumstantial facts. An 
investigation of the composition of a large enough sample of SGB stars is required 
to firmly establish these relations.}

\maketitle

\section{Introduction}

The distribution of stars along the horizontal branch (HB) of globular clusters 
(GC) provides a wealth of information that can be used to understand their
formation and evolution. It is well known that several parameters are required 
to explain the shape of the HBs (van den Bergh 1967; Sandage \& Wildey 1967;
see discussion in Gratton et al. 2010), the two most important being the overall 
metallicity (usually defined by [Fe/H]: Sandage \& Wallerstein 1960; Faulkner 1966), 
and age (see e.g. Dotter et al. 2010). In the recent years, it has become clear 
that star-to-star variations in He content also play an important role (D'Antona
et al. 2002; D'Antona \& Caloi 2004; Carretta et al. 2009; Gratton et al. 2010).
They are related to the phenomenon of multiple populations (Gratton et al. 2004, 2011b; 
Piotto et al. 2008), that is the presence of several generations of stars in 
GCs, the ejecta of a fraction of the stars from the earliest one polluting the 
material from which younger stars formed. Together with He, the abundances of 
several other elements change, including those produced by $p-$capture reactions 
in high-temperature H-burning that can then be used as tracers of this phenomenon. 
Anti-correlations are expected between C and N, O and Na, Mg and Al, depending on 
the temperature at which the H-burning occurred (which in turn is related to the 
mass of the polluters) and then on the timescale of the whole process. The abundance 
of these elements may then be used as a powerful diagnostics of the early phases 
of GC evolution.

Clusters with extended or even discontinuous distribution of stars along the HB 
(see Catelan et al. 1998) may be particularly interesting. In this general frame, 
it should be expected that there is an overall correlation between the colour 
(i.e. temperature) of the stars along the HB and the abundances of He and $p-$capture 
elements (D'Antona \& 
Caloi 2004). While many circumstantial evidence favour this scenario (see e.g. 
Gratton et al. 2010 and references therein), few GCs have been studied with adequate 
data to provide a more direct confirmation. Very recently, the existence of a clear correlation 
between HB morphology and Na-O anticorrelation have been soundly proved for M4 (Marino et al. 2011) and 
by our team for NGC~2808 (Gratton et al. 2011a). However, data on additional GCs
are required because there are other mechanisms that might potentially cause spreads 
in colours of stars along the HB (spread in mass loss, age, metallicity). For instance, such an effect may be produced by
the merging of clusters of different age/chemical composition (van den Bergh 1996; 
Catelan 1997), a phenomenon that might possibly occur within a dwarf galaxy, later 
accreted by the Milky Way (Bellazzini et al. 2008; Bekki \& Yong 2011). Disentangling 
the effects of these various mechanisms is basic to a proper understanding of GC 
formation and evolution.

NGC~1851 is one of the most interesting GCs on this respect. Its HB is very peculiar, 
with a pronounced bimodal distribution of stars: a red HB (RHB), comprising $\sim 54$\% 
of the stars, a blue HB (BHB: $\sim 35$\%) and quite few RR Lyrae variables ($\sim 12$\%: 
Walker 1998; Saviane et al. 1998; Milone et al. 
2009, who obtained a slightly higher fraction of RHB stars)\footnote{The different radial 
distribution of RHB and BHB stars within the cluster 
makes an exact estimate of the relative fractions quite complicate (Saviane et al. 1998; 
Milone et al. 2008, 2009). Here, we use the fractions derived at rather large 
distances from the center (typically in the range 2-10 arcmin) because our spectroscopic 
data mainly sample this region. For comparison, the core and half light
radii of NGC~1851 are 0.09 and 0.51 arcmin, respectively (Harris et al. 1996).}. 
While this might be explained by He abundance variations, in analogy with the 
case of NGC~2808 (D'Antona \& Caloi 2004), the blue HB is actually only slightly brighter 
than the red one, much less than it should be expected for the difference in He 
content required to explain the large spread in colours (see Salaris et al. 2008). 
NGC~1851 shows other important differences from NGC~2808. First, 
the BHB is much shorter, lacking the long blue tail which is very prominent in 
NGC~2808. This agrees with its much fainter total absolute magnitude, because there 
is a good correlation between the maximum temperature of the HB stars and cluster
luminosity (Recio-Blanco et al. 2006). It is an additional indication that 
NGC~1851 lacks very He-rich stars, a fact which is also derived from the 
absence of any discernible split of the MS (Milone et al. 2008). Second, the same 
Milone et al. (2008) showed that the subgiant branch splits into two sequences, 
the faint (f-SGB) one including $34\pm 3$\% of the stars, the remaining 
being on the brighter one (b-SGB). This splitting might be interpreted either as a 
spread of age (about 1 Gyr: Milone et al. 2008) or as the f-SGB being 
overabundant in CNO elements by $\sim 0.3$~dex or more (Cassisi et al. 2008; 
Salaris et al. 2008; Ventura et al. 2009). In the first case, 
the progeny of the f-SGB could be identified with the BHB, while the b-SGB might 
correspond to the RHB; the opposite should hold in the second 
case\footnote{While acknowledging the problem related to number counts,
Salaris et al. (2008) and Ventura et al. (2009) explored the possibility that
a CNO-rich f-SGB coeval with a CNO-normal b-SGB is connected to the BHB. Everything else held
constant, CNO-rich HB's should be redder and not bluer than CNO-poor ones. While
a corresponding variation in He might result in a bluer HB, the variation proposed
by Ventura et al. of $\Delta Y\sim 0.04$ is not enough to justify the proposed
connection. With this hypothesis, the only possibility to reconcile models with
observations would then be to assume that CNO-rich stars experience a larger mass loss along the RGB.}. The
number ratio favours then the first hypothesis, though the abundance analyses
discussed below possibly support the second one. Furthermore, Str\"omgren photometry 
of RGB stars (Grundahl \& Bruntt 2006; Villanova et al. 2009; Carretta et al. 2011a, 2011b) 
shows a peculiar sequence with very 
red $v-y$\ colours that includes some 5-10\% of the stars. This sequence could 
also be explained by an overabundance of CNO elements (Carretta et al. 2011b);
however its relation with the splitting of the SGB and HB is not clear, because 
the fraction of f-SGB stars is much higher than that of stars on the red $v-y$ 
RGB sequence.

Various spectroscopic studies of red giants of NGC~1851 have been carried out in the last
few years. They provided very interesting results, but not yet a final understanding
of this cluster. Yong \& Grundahl (2008) and Yong et al. (2009) studied a few stars with high resolution and high
$S/N$\ spectra. At variance from what is typical of GCs, they found a correlation
between $p-$capture and $n-$capture elements, as well as hints for a spread in CNO
elements. The correlation between $p-$\ and $n-$capture elements has been
confirmed by Villanova et al. (2010) using a slightly larger sample of stars; they
however did not find any spread in CNO elements, possibly because of
uncertainties in the transformation from relative to absolute abundances required
for this determination (see Yong et al. 2011). They also
found that stars on the red $v-y$\ sequence are typically Ba-rich. Carretta et al.
(2010, 2011) considered a much larger sample of red giants, albeit at lower
resolution and S/N. They found a spread in [Fe/H] values larger than typical
in GCs, which they interpreted as due to different populations; this agrees
with what found from narrow band Ca II photometry (Lee et al. 2009). They also 
confirmed the correlation between $n-$\ and $p-$capture elements, and found that Ba-rich
stars are typically more metal-rich than average. The stars on the red
$v-y$\ sequence are indeed Ba-rich, but they found also some Ba-rich stars among
the normal sequence (admittedly, these Ba abundances have rather large errors
being based on a single, less than optimal line).

The overall pattern of abundances of NGC~1851 is clearly peculiar and cannot
easily fit into the scheme adopted for more typical GCs such as M4 or NGC~2808.
The only other cluster where a spread of Fe and a correlation
between $p-$\ and $n-$capture elements have been found is M22, if we leave aside
the two more massive ones, $\omega$~Cen and M54. Developing earlier similar concepts
by  van den Bergh (1997) and Catelan (1998), Carretta et al. (2010, 2011a) proposed that NGC~1851 
is actually the result of the merging of two clusters, each one having their own Na-O
anticorrelation, differing in age (by about 1-2 Gyr) and with a small
difference in [Fe/H]. In their picture, the f-SGB, the Fe and Ba-poor RGB 
population, and the BHB are related to the older cluster; and the b-SGB,
the Fe and Ba-rich RGB, and the RHB to the young one. Since there are however
indications that NGC~1851 hosts also CNO-rich stars, it is possible
that some of the f-SGB stars are actually CNO-rich stars of a younger
population. In this scheme, there should be a complex correlation 
between chemical composition and colours of stars along the RHB, at variance 
with the cases for M~4 and NGC~2808. An explicit study of the chemical
composition of stars along the HB might then help clarifying which is the correct
scenario. In this paper we present the results of such a study. 

The structure of the paper is the following: in Section 2 we present the 
observational data; in Section 3 we explain our analysis methods; in Section 4
we discuss the results of the abundance analysis; conclusions are drawn in Section 5.

\begin{table*}[htb]
\centering
\caption[]{Basic data for program stars}
\begin{scriptsize}
\begin{tabular}{ccccccccc}
\hline
  Star & RA (J2000) & Dec (J2000) & $B$ & $V$ & $K$ & $V_r$& S/N& S/N\\
   &    &         &(mag)&(mag)&(mag)&(km\ts s$^{-1}$)&HR12&HR19\\
\hline
\multicolumn{9}{c}{Blue HB}\\
\hline
~3515 &	5~13~59.619	&	-39~54~21.26	&	16.342	&	16.342	&	15.688	&	322.9	&	26	&	37	\\
13787 &	5~13~42.355	&	-40~09~16.69	&	16.512	&	16.538	&	15.273	&	316.6	&	15	&	30	\\
13858 &	5~13~44.039	&	-40~08~48.36	&	16.312	&	16.276	&	15.765	&	322.6	&	26	&	38	\\
20690 &	5~14~15.237	&	-40~06~28.88	&	16.501	&	16.497	&	15.742	&	323.4	&	17	&	31	\\
21036 &	5~14~15.823	&	-40~06~11.17	&	16.430	&	16.394	&	16.096	&	320.7	&	27	&	35	\\
21285 &	5~14~21.844	&	-40~06~00.42	&	16.723	&	16.737	&	15.678	&	321.3	&	22	&		\\
22551 &	5~13~59.713	&	-40~05~20.51	&	16.864	&	16.912	&			&	320.2	&	20	&	26	\\
26345 &	5~14~12.125	&	-40~04~11.96	&	16.196	&	16.131	&	15.447	&	320.9	&		&	40	\\
26374 &	5~14~17.214	&	-40~04~11.43	&	16.871	&	16.917	&	16.753	&	314.5	&	12	&	27	\\
26686 &	5~14~01.152	&	-40~04~07.72	&	16.461	&	16.458	&	15.019	&	321.4	&	25	&		\\
27792 &	5~14~02.175	&	-40~03~53.79	&	16.945	&	16.979	&			&	330.9	&		&	21	\\
27813 &	5~13~54.243	&	-40~03~53.69	&	16.750	&	16.776	&	15.131	&	328.4	&		&	25	\\
28043 &	5~14~09.589	&	-40~03~50.38	&	16.433	&	16.456	&			&	319.3	&		&	33	\\
28078 &	5~14~14.805	&	-40~03~49.81	&	16.766	&	16.817	&			&	322.0	&	18	&		\\
28346 &	5~14~05.052	&	-40~03~46.82	&	16.267	&	16.187	&			&	322.6	&	25	&	44	\\
29743 &	5~13~57.151	&	-40~03~30.37	&	17.025	&	17.104	&			&	324.9	&		&	22	\\
31860 &	5~14~19.417	&	-40~03~06.26	&	16.765	&	16.824	&			&	318.2	&		&	17	\\
33688 &	5~13~54.712	&	-40~02~46.31	&	16.507	&	16.526	&			&	320.8	&		&	32	\\
34973 &	5~13~55.740	&	-40~02~32.45	&	17.005	&	17.093	&			&	326.5	&	14	&		\\
35774 &	5~14~15.818	&	-40~02~22.70	&	16.471	&	16.468	&	16.166	&	321.5	&		&	34	\\
35868 &	5~14~00.891	&	-40~02~22.26	&	16.075	&	16.057	&	14.014	&	322.7	&	35	&		\\
36193 &	5~13~58.666	&	-40~02~18.30	&	17.175	&	17.282	&			&	309.2	&		&	18	\\
39268 &	5~14~12.776	&	-40~01~43.00	&	16.476	&	16.529	&			&	321.0	&	21	&	31	\\
40227 &	5~14~07.034	&	-40~01~31.43	&	16.826	&	16.902	&			&	328.3	&	16	&	24	\\
40232 &	5~14~20.176	&	-40~01~30.91	&	16.795	&	16.859	&			&	322.5	&	18	&	29	\\
40454 &	5~14~13.145	&	-40~01~28.07	&	16.737	&	16.801	&			&	321.6	&	17	&	30	\\
41678 &	5~14~11.662	&	-40~01~11.22	&	16.792	&	16.886	&	16.294	&	321.8	&	22	&		\\
41796 &	5~14~05.087	&	-40~01~09.76	&	16.436	&	16.491	&			&	321.7	&		&	30	\\
41951 &	5~13~58.932	&	-40~01~07.55	&	16.258	&	16.211	&			&	329.0	&		&	44	\\
43888 &	5~14~08.812	&	-40~00~33.76	&	16.636	&	16.715	&	16.590	&	321.8	&	23	&	31	\\
46632 &	5~14~07.635	&	-39~59~05.32	&	16.708	&	16.787	&	15.395	&	320.5	&	22	&		\\
46902 &	5~14~12.335	&	-39~58~48.16	&	16.202	&	16.098	&	15.739	&	315.7	&	32	&	39	\\
48007 &	5~14~03.922	&	-39~56~57.08	&	16.284	&	16.160	&	15.473	&	320.1	&	32	&	47	\\
52011 &	5~14~33.081	&	-40~03~06.99	&	16.650	&	16.665	&	15.861	&	321.1	&		&	27	\\
52576 &	5~14~53.221	&	-40~01~53.03	&	16.767	&	16.818	&	15.683	&	319.8	&	18	&	20	\\
\hline
\multicolumn{9}{c}{RR Lyrae}\\
\hline
28738 &	5~13~59.846	&	-40~03~41.98	&	16.488 	&	16.122	&	    	&	350  	&		&	51	\\
\hline
\multicolumn{9}{c}{Red HB}\\
\hline
13627 &	5~13~31.981	&	-40~10~16.86	&	16.750	&	16.120	&	14.266	&	318.5	&	26	&	51	\\
20785 &	5~13~56.295	&	-40~06~23.96	&	16.859	&	16.203	&	14.436	&	322.0	&	32	&	62	\\
21988 &	5~13~54.295	&	-40~05~37.13	&	16.911	&	16.241	&	14.381	&	318.4	&	36	&	54	\\
22164 &	5~13~47.857	&	-40~05~31.91	&	16.902	&	16.210	&	14.155	&	319.9	&	39	&	62	\\
22393 &	5~14~07.400	&	-40~05~24.86	&	16.851	&	16.188	&	14.262	&	320.9	&	36	&	54	\\
22548 &	5~14~15.569	&	-40~05~20.03	&	16.855	&	16.196	&	14.258	&	316.6	&	40	&	60	\\
23088 &	5~14~03.290	&	-40~05~07.11	&	16.831	&	16.169	&	14.259	&	318.1	&	38	&	65	\\
\hline
\end{tabular}
\end{scriptsize}
\end{table*} 

\setcounter{table}{0}
 
\begin{table*}[htb]
\centering
\caption[]{Basic data for program stars (Cont.)}
\begin{scriptsize}
\begin{tabular}{ccccccccc}
\hline
  Star & RA (J2000) & Dec (J2000) & $B$ & $V$ & $K$ & $V_r$& S/N& S/N\\
 &  &         &(mag)&(mag)&(mag)&(km\ts s$^{-1}$)&HR12&HR19\\
\hline
23344 &	5~14~09.552	&	-40~05~01.23	&	16.847	&	16.211	&	14.312	&	319.7	&		&	78	\\
24623 &	5~13~59.896	&	-40~04~38.47	&	16.851	&	16.193	&	14.273	&	324.2	&	38	&	58	\\
25243 &	5~14~10.185	&	-40~04~27.63	&	16.797	&	16.149	&	14.144	&	322.4	&	41	&	61	\\
25336 &	5~14~16.616	&	-40~04~26.05	&	16.812	&	16.191	&	14.425	&	319.8	&	34	&	56	\\
25504 &	5~13~54.500	&	-40~04~24.52	&	16.842	&	16.221	&	14.291	&	329.3	&	35	&		\\
25631 &	5~14~11.546	&	-40~04~22.25	&	16.825	&	16.207	&	14.325	&	318.3	&	33	&	85	\\
25715 &	5~13~55.153	&	-40~04~21.52	&	16.845	&	16.182	&	14.194	&	320.3	&		&	69	\\
25793 &	5~13~51.233	&	-40~04~20.51	&	16.781	&	16.173	&	14.383	&	319.8	&		&	78	\\
27604 &	5~14~04.052	&	-40~03~56.00	&	16.849	&	16.217	&	14.552	&	319.0	&	29	&	75	\\
28175 &	5~13~56.135	&	-40~03~49.23	&	16.895	&	16.198	&	14.195	&	319.7	&		&	72	\\
28746 &	5~14~01.579	&	-40~03~41.84	&	16.828	&	16.195	&	14.420	&	319.7	&		&	64	\\
29078 &	5~14~02.880	&	-40~03~37.87	&	16.771	&	16.111	&	14.159	&	317.8	&	36	&		\\
29962 &	5~14~14.102	&	-40~03~27.43	&	16.819	&	16.181	&	14.444	&	317.5	&	32	&		\\
30838 &	5~14~13.425	&	-40~03~17.88	&	16.801	&	16.179	&	14.293	&	321.2	&		&	54	\\
31469 &	5~14~15.520	&	-40~03~10.78	&	16.832	&	16.189	&	14.317	&	327.4	&	36	&	66	\\
31496 &	5~13~57.321	&	-40~03~11.17	&	16.802	&	16.188	&	14.308	&	333.6	&	36	&		\\
31651 &	5~13~57.827	&	-40~03~09.35	&	16.860	&	16.210	&	14.400	&	319.2	&		&	67	\\
31903 &	5~13~56.378	&	-40~03~06.35	&	16.717	&	16.159	&	14.477	&	312.5	&	36	&		\\
32245 &	5~14~13.430	&	-40~03~01.87	&	16.815	&	16.204	&	14.461	&	322.2	&	34	&	61	\\
33196 &	5~14~12.798	&	-40~02~50.98	&	16.808	&	16.185	&			&	313.0	&	38	&	67	\\
34314 &	5~14~13.683	&	-40~02~39.14	&	16.824	&	16.183	&	14.317	&	321.2	&	36	&	68	\\
34386 &	5~13~54.296	&	-40~02~38.91	&	16.746	&	16.084	&	14.176	&	321.6	&	36	&		\\
35789 &	5~14~14.323	&	-40~02~22.58	&	16.756	&	16.111	&	14.432	&	324.1	&	37	&		\\
36599 &	5~14~27.506	&	-40~02~12.94	&	16.822	&	16.199	&	14.140	&	317.6	&	35	&	61	\\
37121 &	5~14~13.327	&	-40~02~07.68	&	16.754	&	16.124	&	14.372	&	322.2	&	34	&		\\
37123 &	5~14~16.085	&	-40~02~07.57	&	16.781	&	16.177	&	14.358	&	321.1	&	34	&		\\
38202 &	5~14~09.724	&	-40~01~55.85	&	16.824	&	16.191	&	14.321	&	320.4	&	41	&	55	\\
39028 &	5~13~49.722	&	-40~01~46.79	&	16.833	&	16.229	&	14.349	&	318.5	&		&	56	\\
39317 &	5~14~05.012	&	-40~01~42.74	&	16.646	&	16.091	&	14.119	&	322.2	&	33	&		\\
39443 &	5~13~56.321	&	-40~01~41.54	&	16.859	&	16.245	&	14.338	&	314.5	&	39	&	61	\\
39832 &	5~13~59.397	&	-40~01~36.81	&	16.748	&	16.118	&	14.319	&	317.0	&	34	&		\\
39984 &	5~13~57.475	&	-40~01~34.81	&	16.830	&	16.215	&	14.456	&	325.2	&	38	&	63	\\
40117 &	5~14~03.457	&	-40~01~32.92	&	16.815	&	16.212	&	14.418	&	317.4	&	21	&	49	\\
40289 &	5~14~06.028	&	-40~01~30.55	&	16.824	&	16.214	&	14.466	&	319.1	&		&	64	\\
40450 &	5~14~02.020	&	-40~01~28.55	&	16.790	&	16.187	&	14.304	&	317.4	&	32	&	72	\\
40767 &	5~13~54.626	&	-40~01~24.45	&	16.820	&	16.214	&	14.313	&	319.1	&	34	&	60	\\
40897 & 5~14~05.974 &   -40~01~22.55    &   16.778  &   16.183  &           &   319.1   &   40  &       \\
41193 &	5~14~22.824	&	-40~01~17.83	&	16.772	&	16.157	&	14.308	&	316.5	&	39	&	72	\\
41381 &	5~14~10.396	&	-40~01~15.82	&	16.792	&	16.222	&	14.599	&	313.1	&	38	&		\\
42849 &	5~14~00.465	&	-40~00~53.43	&	16.755	&	16.169	&	14.420	&	316.7	&	38	&		\\
44554 &	5~14~02.587	&	-40~00~18.74	&	16.825	&	16.227	&	13.147	&	317.2	&	34	&	64	\\
47239 &	5~13~55.946	&	-39~58~23.80	&	16.832	&	16.214	&	14.290	&	316.5	&	30	&	57	\\
47546 &	5~13~59.563	&	-39~57~52.78	&	16.822	&	16.199	&	14.442	&	321.7	&	33	&	60	\\
50923 &	5~14~41.447	&	-40~05~43.73	&	16.699	&	16.129	&	14.483	&	319.9	&	26	&	61	\\
51490 &	5~14~36.982	&	-40~04~16.38	&	16.829	&	16.162	&	14.281	&	316.2	&	40	&	71	\\
51917 &	5~14~35.031	&	-40~03~19.04	&	16.850	&	16.195	&	14.524	&	319.4	&		&	56	\\
54362 &	5~14~42.201	&	-39~56~59.38	&	16.687	&	16.144	&	14.591	&	317.3	&	31	&	48	\\
\hline
\multicolumn{9}{c}{Lower RGB}\\
\hline
20189 &	5~14~04.735	&	-40~07~01.94	&	17.339	&	16.521	&	14.398	&	319.0	&		&	56	\\
21830 &	5~14~06.475	&	-40~05~41.91	&	17.344	&	16.515	&	14.254	&	316.5	&	31	&		\\
25497 &	5~13~59.227	&	-40~04~24.45	&	17.452	&	16.608	&	14.205	&	321.8	&	26	&	55	\\
25799 &	5~14~20.707	&	-40~04~19.53	&	17.344	&	16.493	&	14.226	&	319.8	&		&	56	\\
26532 &	5~14~11.335	&	-40~04~09.43	&	17.293	&	16.472	&	14.301	&	311.9	&	33	&		\\
27085 &	5~14~08.569	&	-40~04~02.44	&	17.306	&	16.528	&	14.346	&	322.3	&	31	&		\\
28445 &	5~13~56.961	&	-40~03~45.67	&	17.567	&	16.753	&	14.450	&	327.2	&	22	&		\\
34604 &	5~13~50.776	&	-40~02~36.82	&	17.487	&	16.701	&	14.328	&	320.6	&		&	49	\\
38250 &	5~14~02.111	&	-40~01~55.53	&	17.441	&	16.628	&	13.282	&	323.2	&	30	&		\\
39992 &	5~14~00.846	&	-40~01~34.62	&	17.303	&	16.490	&	14.378	&	321.1	&		&	48	\\
45111 &	5~14~09.575	&	-40~00~03.16	&	17.487	&	16.720	&	14.324	&	321.8	&		&	45	\\
46657 &	5~14~10.036	&	-39~59~03.68	&	17.316	&	16.521	&	14.236	&	318.5	&	38	&		\\
50876 &	5~14~35.997	&	-40~05~53.11	&	17.512	&	16.680	&	14.431	&	320.7	&	25	&	45	\\
\hline
\end{tabular}
\end{scriptsize}
\label{t:tab1}
\end{table*}

\begin{center}
\begin{figure}
\includegraphics[width=8.8cm]{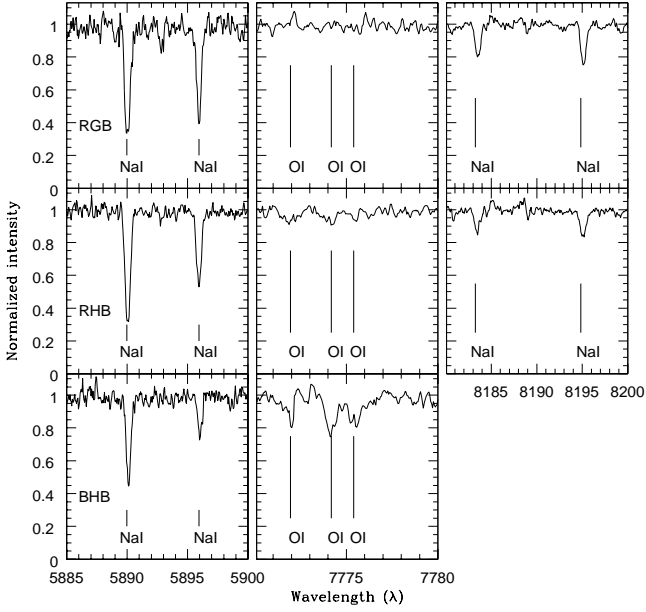}
\caption{Portion of the spectra of an RGB star (\#50876: upper row), 
a RHB (\#22164: middle row), and a BHB star (\#18358: bottom row) }
\label{f:fig1}
\end{figure}
\end{center}

\begin{center}
\begin{figure}
\includegraphics[width=8.8cm]{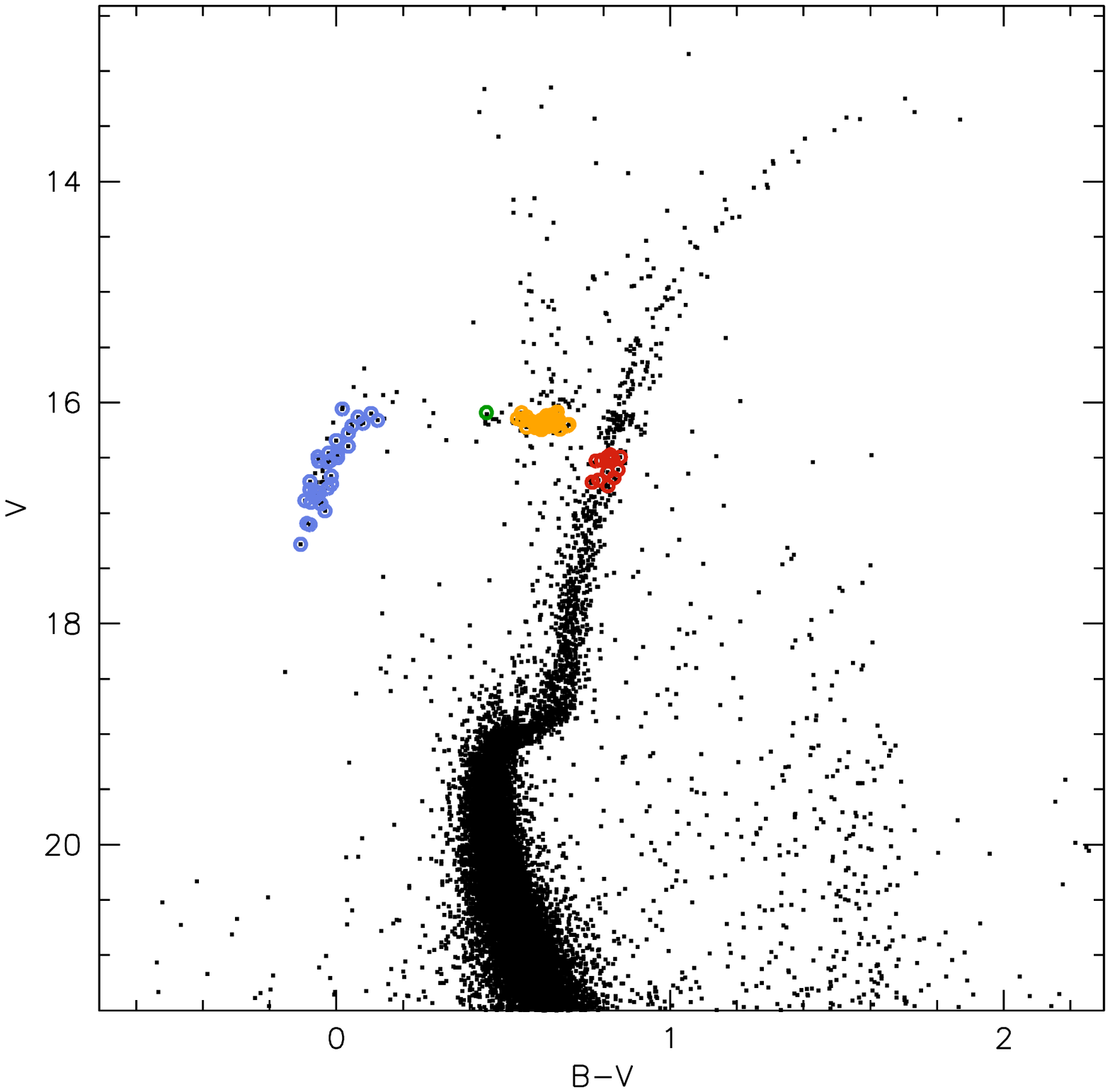}
\caption{Colour-magnitude diagram of the inner
$1\farcm5\le~R~\le10.5$~arcmin region of NGC~1851. Circled stars are those observed
in this paper. The colour code is: blue=BHB; green=RR Lyrae variable;
orange=RHB; red=RGB.}
\label{f:fig2}
\end{figure}
\end{center}

\section{Observation}

We acquired spectra for 35 stars on the BHB, 1 RR Lyrae variable
\footnote{In our original strategy we avoided observing RR Lyrae because it is
difficult to optimize multi-object observations
in service mode for variable stars. However, one RR Lyrae
star was mistakenly included in the sample. Luckily, all spectra turned out to be taken
close to minimum, in the most favourable phase for abundance analysis. This star
was then kept in the analysis.}, 57 stars on the RHB, 
and 13 on the lower RGB (luminosity below the bump)
of NGC~1851 using the GIRAFFE fibre-fed spectrograph at VLT (Pasquini et al. 2004).
All stars were chosen to be free from any companion closer 
than 2 arcsec and brighter than $V+2$~mag, where $V$ is the target magnitude. 
The remaining fibres were used to acquire sky spectra. The median spectra from these 
last fibres were subtracted from those used for the stars. This was of particular 
relevance here, because the observed stars are typically very faint. We used two 
spectral configurations: HR12 (spectral range from 5808 to 6138~\AA) and 
HR19  (from 7728 to 8317~\AA), providing high resolution spectra including 
the strongest features of O~I (the IR triplet at 7771-74~\AA) and Na~I (the 
resonance D doublet at 5890-96~\AA, as well as the subordinate strong doublet at 
8183-94~\AA) accessible from ground and the only ones that might be used to determine 
O and Na abundances without a prohibitively long observing time. A few lines of N, Mg, 
Al, Si, Ca, Fe, and Ba were also included in the selected observing ranges.
Different fibre configurations were used in order to observe a quite large number
of RGB stars with the UVES spectrograph; these observations will be described elsewhere.
We have observations with both gratings only for a subset of stars because a change
in the UVES fibers placements requires a repositioning also of the GIRAFFE fibers to
maximize the number of UVES RGB stars.

Our program was executed in service mode. We obtained a total of 6666~s ($3\times 2222$\,s 
exposures) and 7500~s ($3\times 2500$\,s) of observation with the gratings HR12 
and HR19, respectively. The S/N of the summed spectra for the two gratings is 
typically $\sim 35$ and $\sim 60$, $\sim 22$ and $\sim 32$, and $\sim 30$ and $\sim 50$,
for RHB, BHB, and RGB stars, respectively. The spectra were reduced by the ESO personnel 
using the ESO FLAMES GIRAFFE 
pipeline version 2.8.7. Sky subtraction, combination of individual exposures for 
each star, translation to rest-frame and continuum tracing were performed within 
IRAF \footnote{IRAF is distributed by the National Optical Astronomical 
Observatory, which are operated by the Association of Universities for Research in 
Astronomy, under contract with the National Science Foundation}. Telluric lines 
were removed from the longest wavelength spectra by dividing the average 
spectrum of the warmer BHB stars (those with $T_{\rm eff}>11,500$~K). This 
combined spectrum has a $S/N$\ much higher than those of the individual program 
stars, and was obviously taken with the same airmass, so that the 
excision of the telluric lines turned out to be excellent. Examples of spectra are 
shown in Figure~\ref{f:fig1}.

Figure~\ref{f:fig2} shows the location of the program stars on the colour magnitude
diagram of NGC~1851. Our NGC1851 ground-based photometric catalogue (see  Momany et
al. 2004) consists of $UBV$ observations obtained at the Wide-Field Imager (WFI) 
mounted on the 2.2m  ESO-MPI telescope (La Silla, Chile). Photometric data for the 
program stars 
are listed in Table~\ref{t:tab1}. The $K$\ magnitudes are from the 2MASS point 
source catalogue (Skrutskie et al. 2006), and Str\"omgren photometry is from 
Calamida et al. (2007). 

No information on membership of the program stars to the cluster was available 
prior to the observations. The very high radial velocity of NGC~1851 
($+320.5\pm 0.6$~km\ts s$^{-1}$, Harris 1996) allows to easily rule out field 
stars: a comparison with the Galactic model by Robin et al. (2003) indicates 
a probability of $\sim$2\% to find one field interloper among the whole RHB and RGB 
samples in the same velocity range while the contamination among BHB stars 
is negligible ($P<10^{-5}$). So, stars with radial velocities consistent
with that of the cluster can be quite safely considered cluster members.
On the other hand, all stars observed on the BHB and lower RGB have velocities
consistent with that of the cluster, while two of the candidate RHB stars turned out
to be disk interlopers.
The average radial velocity of the RHB stars is $+319.7\pm 0.5$~km\ts s$^{-1}$\
(r.m.s. scatter of 3.7~km\ts s$^{-1}$),
that of BHB stars is $+321.6\pm 0.7$~km\ts s$^{-1}$\
(r.m.s. scatter of 4.1~km\ts s$^{-1}$), and that of the lower RHB
stars is $+320.3\pm 1.0$~km\ts s$^{-1}$\
(r.m.s. scatter of 3.6~km\ts s$^{-1}$). All these values agree with the value 
listed by Harris (1996). 

Most of the BHB stars of NGC~1851 are cooler than the Grundahl et al. (1999) $u-$jump,
and could be used in our analysis. Only five stars are warmer than this limit.
We did not analyse them because surface abundances for these stars are heavily
influenced by the effects of diffusion and radiation pressure, and then results
are very difficult to be used in a discussion of original abundances. In addition, 
while in general we tried to avoid
observation of variable stars, star \#28738 turned out to be a
known RR Lyrae (V12 of Walker 1998), which was shifted out of the instability
strip in our photometry. Using data by Walker (1998), we
estimated that all our three observations for this star were luckily obtained
close to the minimum of the light curve, and could be used for abundance
analysis. For this star we then adopted a $B-V=0.47$, which is the average colour
at the observed phases, according to Walker (1998).

Some HB stars rotate with velocities up to 
a few tens of km\ts s$^{-1}$\ (Peterson et al. 1995; Behr et al. 2000a, 2000b;
Carney et al. 2008). We checked for fast rotators in our sample
examining the FWHM of the lines, by cross correlating the spectra
with those of templates. No really fast rotator was found in our sample.
The only possible moderate rotator is star \#27792, for which we obtain
a FWHM=38.0~km\ts s$^{-1}$, with respect to typical values of 27~km\ts s$^{-1}$
for the other BHB stars. This might indicate that this star rotates
with $V~\sin{i}\sim 30$~km\ts s$^{-1}$. However, star \#27792 is warmer
than 11,500~K and was not included in our abundance analysis.

\section{Analysis}

\subsection{Atmospheric parameters}

The analysis follows procedures similar to those adopted in the case 
of NGC~2808 (Gratton et al. 2011). However, a few modifications were made, 
so we describe them again.

For the RHB and RGB stars, effective temperatures 
were derived from the $B-V$, $b-y$, and $V-K$\ colours, using the calibration of Alonso 
et al. (1999, with the erratum of Alonso et al. 2001). The colours were dereddened 
using the $E(B-V)$\ values from the updated on-line version of the Harris (1996) 
catalogue and the $E(V-K)/E(B-V)$\ value from Cardelli et al. (1989). The calibrations 
require input values for the metallicity [A/H]. We adopted the value obtained by
Carretta et al. (2010). We assigned weight 4 to the $B-V$\ colours, 5 to the $b-y$, 
and 1 to the $V-K$\ ones, because the program stars are very faint for the 2MASS observations.
The comparison between temperatures from $b-y$ and those from the other two indices
together yields a mean difference of $0\pm 8$~K (r.m.s.=67~K) and $5\pm 20$~K
(r.m.s.=71~K) for the RHB and RGB stars, respectively. For the RR Lyrae star \#28738=V12,
we could only use the $B-V$\ value from Walker (1998).

For the blue HB stars, we started from the $(B-V) - T_{\rm eff}$\ calibration by 
Kurucz\footnote{See kurucz.harvard.edu.}, as in NGC~2808. Infrared colours from 2MASS 
are not reliable for these faint stars. Furthermore, $B-V$\ colours saturate, so that errors in 
individual temperature values become very large. We then derived effective temperatures 
from $B-V$ for individual stars ($T_{\rm eff}(B-V)$), but then fitted a quadratic 
relation between these temperatures and the $V$ magnitudes, and extracted a temperature
($T_{\rm eff}(V)$) entering the $V$ magnitude into this relation. We repeated this
procedure using $u-y$\ colours from Str\"omgren photometry rather than $V$, obtaining a 
second value for the temperature
($T_{\rm eff}(u-y)$). The adopted $T_{\rm eff}$'s are the average of $T_{\rm eff}(V)$ and 
$T_{\rm eff}(u-y)$ for individual stars obtained from these relation. Since they are 
ultimately calibrated against $T_{\rm eff}(B-V)$, these temperatures are on the same 
scale used for NGC~2808, but they have much smaller internal errors. Note that two stars
(\#35868 and \#46902) are clearly brighter than the mean line for the BHB; they are likely
stars evolved off the HB. For these stars, $T_{\rm eff}(V)$\ was not considered.
On average, $T_{\rm eff}(V)-T_{\rm eff}(u-y)=33\pm 52$~K, with an r.m.s. of 280 K.

Given these comparisons, we assumed errors of 50, 50, and 200~K as representative values 
for the internal errors in the temperatures for RHB, RGB, and BHB stars respectively. 
Systematic errors due to scale errors or incorrect parameters for the cluster are 
likely larger. We come back later on their potential impact.

The surface gravities were obtained from the masses, luminosities, and effective temperatures. For 
the masses, we adopted values of 1.00, 0.657, and 0.575~$M_\odot$\ for stars on the RGB,
RHB, and BHB (see Gratton et al. 2010 and Sect. 4.5 for discussions of
values adequate for the different sequences). The bolometric corrections
were obtained using calibrations consistent with those used for the effective temperatures (Alonso et
al. 1999 for the red giant and RHB stars, and Kurucz for the BHB
stars). The adopted distance modulus has been taken from Harris' catalogue.

Errors in gravities are small. The assumption about masses is likely correct within 10\%, 
while those on the effective temperature and luminosity cause errors in
gravities not larger than $\sim 2$\% for the RHB stars and red giants, and $\sim 8$\%
for the BHB stars. The error in gravities is then not larger than 0.05~dex for the
cool stars and 0.10~dex for the warm ones.

The same metal abundance of [A/H]=-1.2 and microturbulence velocity $\xi_\mu$\ of
2.0~km\ts s$^{-1}$\ were adopted for all RHB stars, and 1.5~km\ts s$^{-1}$\ for the
RGB ones. The metal abundance is similar to the average value of our Fe abundances 
for the RHB stars: [Fe/H]=$-1.14\pm 0.01$ (r.m.s.=0.064 dex). Note that uncertainties 
in the Fe abundances are much larger than represented by this tiny error bar, which is 
simply the standard deviation of the mean value. Our average Fe abundance is very close to that
derived for RGB stars by Carretta et al. (2010, 2011a: [Fe/H]=-1.16). For the RGB stars we 
obtained a slightly lower value of [Fe/H]=$-1.18\pm 0.03$ with a larger r.m.s. of 0.11 dex,
which is not surprising because on average we measured fewer lines due to smaller spectral 
coverage and lower S/N of the spectra. 

The microturbulence velocity $\xi_\mu$\ is not well constrained by 
our data, because the Fe~I lines have only a moderate range in equivalent widths. 
Practically, only the 6065~\AA\ line is saturated enough to really constrain it,
so that its value is sensitive to the atomic parameters (oscillator strength and
damping constant) we adopted for this line. The {\it gf} for this line is from VALD 
database (Kupka et al. 2000)\footnote{See URL vald.astro.univie.ac.at}, 
the damping constant is from Barklem et al. (2000).
Taken at face value, the Fe I lines yield a low value for $\xi_\mu=1.3$~km\ts s$^{-1}$. 
On the other hand, some lines of other elements are quite strong. For instance,
the Ca I lines at 5857 and 6122~\AA\ are well on the flat part of the
curve of growth (see Sect. 3.2 for a discussion of the parameters we adopted for
these lines). The adoption of the low value of the microturbulence velocity
indicated by the Fe~I lines would yield large overabundances of Ca (on average
[Ca/Fe]$\sim 0.9$), inconsistent with the value derived from giants 
([Ca/Fe]=$0.30\pm 0.02$: Carretta et al. 2011a, 2011b), 
which is a typical value for metal-poor stars. The value of $\xi_\mu$\ we 
adopted is a compromise, producing only a moderate trend of the Fe abundances
with EW and a more acceptable average value of [Ca/Fe]=0.48. However, these
comparisons indicate that our values of the microturbulent velocities have
rather large systematic error bars attached, which we estimate at $\pm 0.5$~km\ts s$^{-1}$.
We note that the value we adopted is in the middle of the range usually found in previous analysis 
of BHB stars (Lambert et al. 1992; Behr et al. 1999, 2000b; Kinman et al. 2000; 
Fabbian et al. 2005; Villanova et al. 2009; Marino et al. 2011).

Two Fe~II lines (at 5991.38 and 6084.10~\AA) could be measured in RHB spectra
and only the first one in RGB ones. Abundances derived from these lines are in fair agreement with 
those obtained from the Fe~I lines: on average we obtained [Fe/H]=$-1.20\pm 0.01$ and
$-1.23\pm 0.08$\ for RHB and RGB stars respectively. This supports the choice of the atmospheric
parameters adopted throughout this paper.
  
\begin{table*}[htb]
\centering
\caption[]{Atmospheric parameters and Fe abundances}
\begin{scriptsize}
\begin{tabular}{ccccccccc}
\hline
Star &$T_{\rm eff}$&$\log{g}$&       &$[$Fe/H$]_I$& &  &$[$Fe/H$]_{II}$& \\
     &(K)          & (dex)   & lines & $<>$       & rms &lines & $<>$          & rms\\
\hline
\multicolumn{9}{c}{Blue HB}\\
\hline
3515  &	~9343 &	3.35	&		&		&		&		&		&		\\
13787 &	10025 &	3.50	&		&		&		&		&		&		\\
13858 &	~9142 &	3.34	&		&		&		&		&		&		\\
20690 &	~9981 &	3.54	&		&		&		&		&		&		\\
21036 &	~9459 &	3.44	&		&		&		&		&		&		\\
21285 &	10906 &	3.76	&		&		&		&		&		&		\\
22551 &	11517 &	3.83	&		&		&		&		&		&		\\
26345 &	~8767 &	3.23	&		&		&		&		&		&		\\
26374 &	11574 &	3.85	&		&		&		&		&		&		\\
26686 &	~9721 &	3.47	&		&		&		&		&		&		\\
27792 &	11986 &	3.67	&		&		&		&		&		&		\\
27813 &	11083 &	3.77	&		&		&		&		&		&		\\
28043 &	~9680 &	3.42	&		&		&		&		&		&		\\
28078 &	11094 &	3.72	&		&		&		&		&		&		\\
28346 &	~8789 &	3.26	&		&		&		&		&		&		\\
29743 & 12694 & 3.85	&		&		&		&		&		&		\\
31860 &	11695 &	3.79	&		&		&		&		&		&		\\
33688 &	~9976 &	3.51	&		&		&		&		&		&		\\
34973 & 12562 & 3.87	&		&		&		&		&		&		\\
35774 &	~9716 &	3.48	&		&		&		&		&		&		\\
35868 &	~9072 &	3.22	&		&		&		&		&		&		\\
36193 & 13670 & 4.03	&		&		&		&		&		&		\\
39268 &	~9980 &	3.41	&		&		&		&		&		&		\\
40227 &	11368 &	3.71	&		&		&		&		&		&		\\
40232 &	11316 &	3.73	&		&		&		&		&		&		\\
40454 &	10957 &	3.65	&		&		&		&		&		&		\\
41678 &	11762 &	3.68	&		&		&		&		&		&		\\
41796 &	10006 &	3.40	&		&		&		&		&		&		\\
41951 &	~8861 &	3.27	&		&		&		&		&		&		\\
43888 &	10647 &	3.50	&		&		&		&		&		&		\\
46632 &	10980 &	3.59	&		&		&		&		&		&		\\
46902 &	~8902 &	3.26	&		&		&		&		&		&		\\
48007 &	~8751 &	3.25	&		&		&		&		&		&		\\
52011 &	10730 &	3.70	&		&		&		&		&		&		\\
52576 &	11216 &	3.74	&		&		&		&		&		&		\\
\hline
\multicolumn{9}{c}{RR Lyrae}\\
\hline
28738 & ~6014 & 2.70  &  3  &   -1.22 & 0.28    &		&		&		\\
\hline
\multicolumn{9}{c}{Red HB}\\
\hline
13627 &	5441 &	2.44	&	16	&	-1.06	&	0.13	&	2	&	-0.99	&	0.21	\\
20785 &	5405 &	2.46	&	16	&	-1.10	&	0.13	&	2	&	-1.06	&	0.10	\\
21988 &	5348 &	2.46	&	16	&	-1.17	&	0.20	&	2	&	-1.25	&	0.05	\\
22164 &	5306 &	2.42	&	16	&	-1.29	&	0.22	&	1	&	-1.14	&			\\
22393 &	5384 &	2.45	&	16	&	-1.10	&	0.17	&	2	&	-1.14	&	0.02	\\
22548 &	5412 &	2.46	&	16	&	-1.14	&	0.18	&	2	&	-1.21	&	0.06	\\
23088 &	5369 &	2.43	&	16	&	-1.21	&	0.14	&	2	&	-1.27	&	0.14	\\
\hline
\end{tabular}
\end{scriptsize}
\end{table*} 

\setcounter{table}{1}
 
\begin{table*}[htb]
\centering
\caption[]{Atmospheric parameters and Fe abundances (cont.)}
\begin{scriptsize}
\begin{tabular}{ccccccccc}
\hline
Star &$T_{\rm eff}$&$\log{g}$&       &$[$Fe/H$]_I$& &  &$[$Fe/H$]_{II}$& \\
     &(K)          & (dex)   & lines & $<>$       & rms &lines & $<>$          & rms\\
\hline
23344 &	5445 &	2.48	&	7	&	-1.16	&	0.08	&		&		&		\\
24623 &	5415 &	2.46	&	16	&	-1.25	&	0.16	&	2	&	-1.36	&	0.05	\\
25243 &	5394 &	2.44	&	14	&	-1.14	&	0.13	&	2	&	-1.24	&	0.14	\\
25336 &	5519 &	2.50	&	16	&	-1.07	&	0.19	&	2	&	-1.24	&	0.23	\\
25504 &	5411 &	2.48	&	9	&	-1.22	&	0.18	&	2	&	-1.12	&	0.09	\\
25631 &	5500 &	2.50	&	16	&	-1.08	&	0.18	&	2	&	-1.16	&	0.03	\\
25715 &	5386 &	2.44	&	7	&	-1.22	&	0.07	&		&			&			\\
25793 &	5537 &	2.50	&	7	&	-1.17	&	0.11	&		&			&			\\
27604 &	5471 &	2.50	&	16	&	-1.12	&	0.18	&	2	&	-1.16	&	0.00	\\
28175 &	5289 &	2.41	&	7	&	-1.13	&	0.17	&		&			&			\\
28746 &	5424 &	2.47	&	7	&	-1.18	&	0.13	&		&			&			\\
29078 &	5341 &	2.40	&	9	&	-1.13	&	0.20	&	2	&	-1.26	&	0.06	\\
29962 &	5464 &	2.48	&	9	&	-1.01	&	0.14	&	2	&	-1.22	&	0.02	\\
30838 &	5479 &	2.48	&	7	&	-1.07	&	0.14	&		&			&			\\
31469 &	5417 &	2.46	&	16	&	-1.14	&	0.22	&	2	&	-1.41	&	0.24	\\
31496 &	5482 &	2.49	&	9	&	-1.14	&	0.23	&	2	&	-1.17	&	0.10	\\
31651 &	5401 &	2.47	&	7	&	-1.15	&	0.14	&		&			&			\\
31903 &	5691 &	2.55	&	9	&	-1.13	&	0.23	&		&			&			\\
32245 &	5497 &	2.50	&	16	&	-1.06	&	0.20	&	2	&	-1.32	&	0.00	\\
33196 &	5455 &	2.48	&	16	&	-1.05	&	0.18	&	2	&	-1.12	&	0.09	\\
34314 &	5408 &	2.46	&	16	&	-1.10	&	0.19	&	2	&	-1.23	&	0.08	\\
34386 &	5339 &	2.39	&	9	&	-1.18	&	0.06	&	2	&	-1.13	&	0.02	\\
35789 &	5428 &	2.44	&	9	&	-1.10	&	0.16	&	2	&	-1.29	&	0.04	\\
36599 &	5436 &	2.47	&	16	&	-1.13	&	0.16	&	2	&	-1.14	&	0.05	\\
37121 &	5452 &	2.45	&	9	&	-1.11	&	0.17	&	2	&	-1.21	&	0.04	\\
37123 &	5528 &	2.50	&	9	&	-1.11	&	0.20	&	2	&	-1.30	&	0.14	\\
38202 &	5414 &	2.46	&	16	&	-1.20	&	0.12	&	2	&	-1.33	&	0.16	\\
39028 &	5477 &	2.50	&	7	&	-1.12	&	0.12	&		&			&			\\
39317 &	5654 &	2.51	&	9	&	-1.04	&	0.20	&	2	&	-1.19	&	0.10	\\
39443 &	5456 &	2.50	&	15	&	-1.10	&	0.12	&	2	&	-1.18	&	0.25	\\
39832 &	5405 &	2.43	&	9	&	-1.07	&	0.19	&	2	&	-1.28	&	0.02	\\
39984 &	5487 &	2.50	&	16	&	-1.07	&	0.13	&	2	&	-0.98	&	0.08	\\
40117 &	5472 &	2.50	&	15	&	-1.18	&	0.26	&	1	&	-0.98	&			\\
40289 &	5459 &	2.49	&	7	&	-1.20	&	0.19	&		&			&			\\
40450 &	5483 &	2.49	&	15	&	-1.14	&	0.16	&	2	&	-1.14	&	0.04	\\
40767 &	5474 &	2.50	&	16	&	-1.20	&	0.18	&	2	&	-1.31	&	0.06	\\
40897 &	5569 &	2.52	&	9	&	-1.11	&	0.18	&	2	&	-1.18	&	0.02	\\
41193 &	5477 &	2.48	&	16	&	-1.12	&	0.18	&	2	&	-1.10	&	0.21	\\
41381 &	5603 &	2.55	&	8	&	-1.15	&	0.22	&	2	&	-1.14	&	0.01	\\
42849 &	5538 &	2.50	&	9	&	-1.22	&	0.13	&	2	&	-1.34	&	0.09	\\
44554 &	5520 &	2.52	&	15	&	-1.19	&	0.13	&	1	&	-1.42	&			\\
47239 &	5431 &	2.48	&	16	&	-1.18	&	0.15	&	2	&	-1.25	&	0.00	\\
47546 &	5508 &	2.50	&	16	&	-1.13	&	0.20	&	2	&	-1.15	&	0.01	\\
50923 &	5653 &	2.53	&	14	&	-1.18	&	0.25	&	2	&	-1.27	&	0.25	\\
51490 &	5345 &	2.42	&	16	&	-1.18	&	0.13	&	2	&	-1.14	&	0.12	\\
51917 &	5431 &	2.47	&	7	&	-1.36	&	0.18	&		&			&			\\
54362 &	5755 &	2.57	&	15	&	-1.15	&	0.19	&	2	&	-1.31	&	0.15	\\
\hline
\multicolumn{9}{c}{RGB}\\
\hline
20189 &	4957 &	2.58	&	5	&	-1.35	&	0.16	&		&			&			\\
21830 &	4928 &	2.56	&	8	&	-1.15	&	0.16	&	1	&	-1.55	&			\\
25497 &	4909 &	2.58	&	10	&	-1.13	&	0.11	&	1	&	-1.28	&			\\
25799 &	4932 &	2.55	&	7	&	-1.17	&	0.15	&		&			&			\\
26532 &	4948 &	2.56	&	10	&	-1.09	&	0.12	&	1	&	-1.04	&			\\
27085 &	5015 &	2.61	&	10	&	-1.02	&	0.13	&	1	&	-1.16	&			\\
28445 &	4965 &	2.67	&	9	&	-1.10	&	0.12	&		&			&			\\
34604 &	4983 &	2.66	&	7	&	-1.28	&	0.19	&		&			&			\\
38250 &	4936 &	2.62	&	10	&	-1.09	&	0.17	&	1	&	-1.08	&			\\
39992 &	4937 &	2.56	&	5	&	-1.39	&	0.14	&		&			&			\\
45111 &	5023 &	2.70	&	6	&	-1.24	&	0.23	&		&			&			\\
46657 &	4946 &	2.58	&	10	&	-1.14	&	0.17	&	1	&	-1.13	&			\\
50876 &	4908 &	2.62	&	14	&	-1.25	&	0.23	&		&			&			\\
\hline
\end{tabular}
\end{scriptsize}
\label{t:tab2}
\end{table*}

\begin{table*}[htb]
\centering
\caption[]{Abundances of other elements}
\begin{scriptsize}
\begin{tabular}{ccccccccccc}
\hline
Star	&	[N/Fe]	&	[O/Fe]	&	[Na/Fe]	&	[Mg/Fe]	&	[Mg/Fe]	&	[Al/Fe]	&	[Si/Fe]	&	[Ca/Fe]	&	[Mn/Fe]	&	[Ba/Fe]	\\
	&		&		&			&	I	&	II	&		&		&		&		&		\\
\hline
\multicolumn{11}{c}{Blue HB}\\
\hline
3515	&	1.21	&	~0.24	&	~0.44	&		&		&		&		&		&		&		\\
13787	&	1.19	&	~0.08	&	~0.11	&		&	0.68&		&		&		&		&		\\
13858	&	1.05	&	~0.12	&	~0.79	&		&	0.81&		&		&		&		&		\\
20690	&	1.31	&	-0.50	&	~0.72	&		&	0.53&		&		&		&		&		\\
21036	&	1.32	&	~0.56	&	~0.45	&		&	0.61&		&		&	    &		&		\\
21285	&			&			&	~0.31	&		&		&		&		&		&		&		\\
22551	&	1.13	&	-0.12	&	$<$0.81	&		&	0.45&		&		&		&		&		\\
26345	&	1.12	&	~0.20	&	~0.30	&		&	0.72&		&		&		&		&		\\
26374	&	1.44	&	~0.02	&	$<$0.81	&		&	0.71&		&		&		&		&		\\
26686	&			&			&	$<$0.02	&		&		&		&		&		&		&		\\
27813	&	1.37	&	-0.02	&			&		&	0.76&		&		&		&		&		\\
28043	&	1.07	&	~0.26	&			&		&	0.62&		&		&		&		&		\\
28078	&			&			&	~1.17	&		&		&		&		&		&		&		\\
28346	&	0.91	&	-0.10	&	~0.60	&		&	0.51&		&		&		&		&		\\
33688	&	1.18	&	-0.04	&			&		&	0.34&		&		&		&		&		\\
35774	&	1.15	&	~0.18	&			&		&	0.42&		&		&		&		&		\\
35868	&			&			&	~0.33	&		&		&		&		&		&		&		\\
39268	&	1.18	&	-0.01	&	~0.69	&		&	0.80&		&		&		&		&		\\
40227	&	1.23	&	~0.06	&	~1.48	&		&	0.52&		&		&		&		&		\\
40232	&	1.11	&	-0.27	&	$<$0.76	&		&	0.64&		&		&		&		&		\\
40454	&	1.25	&	-0.03	&	~0.73	&		&	0.07&		&		&		&		&		\\
41796	&	1.10	&	-0.37	&			&		&		&		&		&		&		&		\\
41951	&	1.12	&	~0.15	&			&		&	0.55&		&		&		&		&		\\
43888	&	1.04	&	-0.12	&	$<$0.56	&		&	0.58&		&		&		&		&		\\
46632	&			&			&	~0.56	&		&		&		&		&		&		&		\\
46902	&			&	-1.20	&	-0.56	&		&		&		&		&		&		&		\\
48007	&	0.81	&	~0.31	&	~0.99	&		&	0.12&		&		&	    &		&		\\
52011	&	1.31	&	~0.37	&			&		&	1.10&		&		&		&		&		\\
52576	&			&			&	$<$0.71	&		&		&		&		&		&		&		\\
\hline
\multicolumn{11}{c}{RR Lyrae}\\
\hline
28738 & 1.03  &  0.17 & 0.51    &$<$-0.34&		&		&		&		&		&		\\
\hline
\multicolumn{11}{c}{Red HB}\\
\hline
13627	&		&	0.26	&	-0.07	&	0.41	&		&		&	0.16	&	0.64	&		&	0.23\\
20785	&		&	0.37	&	-0.10	&	0.41	&		&		&	0.16	&	0.45	&		&	0.16\\
21988	&		&	0.15	&	-0.17	&	0.34	&		&	~0.12&	0.23	&	0.20	&		&	0.22\\
22164	&		&	0.27	&	-0.10	&	0.25	&		&		&	0.18	&	0.20	&		&	0.13\\
22393	&		&	0.16	&	-0.20	&	0.35	&		&		&	0.30	&	0.30	&		&  -0.17\\
22548	&		&	0.39	&	-0.07	&	0.24	&		&		&	0.21	&	0.17	&		&  -0.10\\
23088	&		&	0.46	&	-0.21	&	0.40	&		&	-0.16&	0.19	&	0.16	&		&  -0.02\\
\hline
\end{tabular}
\end{scriptsize}
\end{table*} 

\setcounter{table}{2}

\begin{table*}[htb]
\centering
\caption[]{Abundances of other elements (cont.)}
\begin{scriptsize}
\begin{tabular}{ccccccccccc}
\hline
Star	&[N/Fe]	&	[O/Fe]	&	[Na/Fe]	&	[Mg/Fe]	&[Mg/Fe]&[Al/Fe]&	[Si/Fe]	&	[Ca/Fe]	&[Mn/Fe]&[Ba/Fe]\\
		&		&			&			&	I		&II		&		&			&			&		&		\\
\hline
23344	&		&	0.36	&	0.03	&	0.38	&		&		&	0.09	&			&		&		\\
24623	&		&	0.46	&  -0.08	&	0.36	&		&		&	0.04	&	0.25	&		&  -0.13\\
25243	&		&	0.38	&  -0.03	&	0.38	&		&		&	0.25	&	0.44	&		&	0.12\\
25336	&		&	0.29	&  -0.01	&	0.51	&		&		&	0.01	&	0.51	&		&	0.01\\
25504	&		&			&  -0.17	&			&		&		&	0.10	&	0.55	&		&	0.32\\
25631	&		&	0.14	&	0.16	&	0.48	&		&		&	0.24	&	0.37	&		&	0.68\\
25715	&		&	0.37	&  -0.17	&	0.31	&		&		&	0.09	&			&		&		\\
25793	&		&	0.39	&	0.46	&	0.43	&		&	0.07&	0.23	&			&		&		\\
27604	&		&	0.31	&	0.08	&	0.42	&		&		&	0.26	&	0.71	&		&	0.22\\
28175	&		&	0.47	&	0.03	&	0.39	&		&		&	0.25	&			&		&		\\
28746	&		&	0.39	&  -0.18	&			&		&		&	0.13	&			&		&		\\
29078	&		&			&	0.10	&			&		&		&	0.14	&	0.52	&		&	0.29\\
29962	&		&			&  -0.03	&			&		&		&	0.07	&	0.40	&		&	0.38\\
30838	&		&	0.19	&	0.51	&	0.43	&		&		&	0.16	&			&		&		\\
31469	&		&	0.26	&	0.01	&	0.32	&		&		&	0.26	&	0.51	&		&	0.34\\
31496	&		&			&	0.06	&			&		&		&	0.23	&	0.28	&		&	0.05\\
31651	&		&	0.34	&  -0.09	&	0.43	&		&		&	0.19	&			&		&		\\
31903	&		&			&	0.93	&			&		&		&	0.23	&	0.64	&		&	1.19\\
32245	&		&	0.54	&	0.07	&	0.37	&		&		&	0.18	&	0.25	&		&	0.25\\
33196	&		&	0.31	&   0.00	&	0.40	&		&		&	0.21	&	0.26	&		&	0.13\\
34314	&		&	0.47	&  -0.04	&	0.40	&		&		&	0.11	&	0.55	&		&	0.26\\
34386	&		&			&	0.02	&			&		&		&	0.32	&	0.39	&		&	0.08\\
35789	&		&			&  -0.10	&			&		&		&	0.25	&	0.42	&		&	0.34\\
36599	&		&	0.33	&	0.04	&	0.43	&		&		&	0.21	&	0.30	&		&	0.13\\
37121	&		&			&	0.40	&			&		&		&	0.39	&	0.52	&		&	1.18\\
37123	&		&			&	0.42	&			&		&		&	0.39	&	0.32	&		&	1.19\\
38202	&		&	0.07	&  -0.15	&	0.38	&		&		&	0.07	&	0.52	&		&  -0.08\\
39028	&		&	0.25	&  -0.04	&	0.42	&		&		&	0.19	&			&		&		\\
39317	&		&			&	0.44	&			&		&		&	0.37	&	0.47	&		&	0.01\\
39443	&		&	0.30	&  -0.04	&	0.47	&		&		&	0.19	&	0.44	&		&	0.28\\
39832	&		&			&	0.04	&			&		&		&	0.22	&	0.62	&		&	0.15\\
39984	&		&	0.32	&	0.11	&	0.33	&		&		&	0.27	&	0.38	&		&	0.34\\
40117	&		&	0.32	&	0.21	&	0.49	&		&		&	0.22	&	0.64	&		&  -0.02\\
40289	&		&	0.28	&	0.03	&	0.42	&		&		&	0.12	&			&		&		\\
40450	&		&	0.46	&  -0.17	&	0.43	&		&		&	0.28	&	0.42	&		&	0.22\\
40767	&		&	0.17	&  -0.11	&	0.31	&		&		&	0.11	&	0.50	&		&	0.03\\
40897	&		&			&  -0.01	&			&		&		&	0.25	&	0.51	&		&	0.17\\
41193	&		&	0.37	&  -0.08	&	0.45	&		&		&	0.00	&	0.37	&		&	0.06\\
41381	&		&			&	0.25	&			&		&		&	0.13	&	0.54	&		&	1.10\\
42849	&		&			&	0.20	&			&		&		&	0.30	&	0.23	&		&	0.26\\
44554	&		&	0.31	&	0.14	&	0.34	&		&		&	0.16	&	0.22	&		&	0.63\\
47239	&		&	0.38	&  -0.13	&	0.38	&		&		&	0.31	&	0.42	&		&	0.06\\
47546	&		&	0.32	&	0.51	&	0.48	&		&		&	0.12	&	0.51	&		&	1.10\\
50923	&		&	0.12	&	0.48	&	0.36	&		&		&	0.01	&	0.60	&		&  -0.36\\
51490	&		&	0.35	&  -0.20	&	0.37	&		&		&	0.29	&	0.24	&		&  -0.10\\
51917	&		&	0.25	&  -0.23	&	0.35	&		&		&	0.04	&			&		&		\\
54362	&$<1.55$&	0.00	&	0.43	&	0.48	&		&		&	0.28	&	0.46	&		&	0.93\\
\hline
\multicolumn{11}{c}{RGB}\\
\hline
20189	&		&	0.48	&	0.06	&	0.31	&		&		&	0.12	&			&		&		\\
21830	&		&			&	0.12	&			&		&		&	0.33	&	0.26	&	-0.47&	0.02\\
25497	&		&			&	0.02	&			&		&		&	0.07	&	0.22	&	-0.50&	0.02\\
25799	&		&	0.72	&  -0.19	&	0.27	&		&		&  -0.07	&			&		&		\\
26532	&		&			&	0.25	&			&		&		&	0.44	&	0.26	&	-0.41&	0.73\\
27085	&		&			&	0.16	&			&		&		&	0.24	&	0.21	&	-0.26&	0.80\\
28445	&		&			&	0.02	&			&		&		&	0.38	&	0.33	&	-0.36&	1.24\\
34604	&		&	0.50	&	0.05	&	0.42	&		&		&	0.18	&			&		&		\\
38250	&		&			&  -0.19	&			&		&		&	0.29	&	0.27	&	-0.36&	0.48\\
39992	&		&	0.22	&  -0.02	&	0.39	&		&		&	0.10	&			&		&		\\
45111	&		&			&  -0.11	&	0.33	&		&		&	0.21	&			&		&		\\
46657	&		&			&  -0.25	&			&		&		&	0.08	&	0.15	&	-0.46&	0.01\\
50876	&		&	0.30	&	0.05	&	0.43	&		&		&	0.04	&  -0.01	&	-0.64&	0.69\\
\hline
\end{tabular}
\end{scriptsize}
\label{t:tab3}
\end{table*}

Table~\ref{t:tab2} lists the effective temperatures $T_{\rm eff}$\ and
surface gravities $\log{g}$ we used in the analysis of the program stars, 
as well as the abundances for Fe~I, Fe~II. Table~\ref{t:tab3} 
gives the abundances for N~I, O~I, Na~I, Mg~I, Mg~II, Si~I, Ca~I, and Ba~II. Abundances were
estimated from equivalent widths. The analysis is very similar to that described
in Gratton et al. (2011a) for NGC~2808. In the following section
we give details for a few elements, outlining what was changed from that
paper.

\subsection{Analysis for individual elements}

Nitrogen: N abundances were derived only for BHB stars (upper limits were obtained for
the cooler stars). They are based on the two high excitation lines at 8216.3 and 
8242.4~\AA. The first one has been used in the recent analysis of the solar N 
abundance by Caffau et al. (2009), who obtained a N abundance of $\log{n(N)}=7.85$\ 
from this line (for the 1-D LTE analysis, which is within 0.01 dex from the value they 
obtain from the 3-D NLTE one), very close to their recommended value of 
$\log{n(N)}=7.86$. We use the VALD $\log{gf}$\ for this line, which is 0.13 dex 
lower than the NIST one; with this value, we obtain a solar N abundance of 
$\log{n(N)}=7.99$ using the Kurucz 1-D solar spectrum, which is consistent with the 
difference in the adopted $gf$'s. We conclude that these lines yield abundances 
consistent with the best estimate of N abundances for the Sun.

On the other hand, non-LTE corrections are likely not negligible for these lines
for BHB stars. Statistical equilibrium calculations for N in population I A-type 
stars have been presented by Przybilla \& Butler (2001), who also made comparisons 
with previous determinations. The stars considered by these authors bracket the 
surface gravity and line strength range of the stars studied here, though they are 
more metal-rich. While they do not provide non-LTE corrections for the 
two lines considered in this analysis, they provide data for many lines of the same 
lower and close upper levels, which are most likely very close to those appropriate
for the lines we could measure. The non-LTE abundance corrections they obtained for 
these lines are nearly proportional to the EWs, being well reproduced by the same 
relation $\Delta$[N/Fe]=-0.0036~$\times$~EW for all stars in their sample. We then adopted the corrections 
given by this relation to the abundances we derived from the LTE analysis. The 
corrections are quite uniform, with a mean value of -0.28~dex. Although it is clear 
that this is a rough procedure that may bring some additional uncertainty, we deem 
unlikely that these corrections are in error by more than half this value. After 
this correction, the N abundances are very uniform among the BHB stars, 
with an average value of [N/Fe]=$1.16\pm 0.15$. The error bar is here the r.m.s. 
scatter of individual values and agrees with internal errors. We plotted these N 
abundances against various quantities
(including O and Na abundances); we only found a small trend for increasing N abundances
with effective temperature, with stars with $T_{\rm eff}<9000$~K having N abundances
some 0.1-0.2~dex below the average, and those with $T_{\rm eff}>11000$~K with N
abundances higher than average by a similar amount. This small trend might either be 
real (warmer stars on the HB might indeed be expected to be more N-rich), or an 
artifact of the analysis, since the trend is at the level where we expect possible
systematic errors.

We looked for the N lines in the spectra of the RHB stars. The line at 8216.3~\AA\ 
is in a difficult region, with strong telluric lines that must be subtracted with care. 
We looked for but did not detect the 8242.6~\AA\ line in the summed spectrum of the RHB 
stars; we may set an upper limit of EW$<3$~m\AA, which yields [N/Fe]$<1.1$. We also 
looked for this line in individual spectra but we did not detect it in any. 
For instance, the robust upper limit of 15~m\AA\ we get for the warm, Ba-rich star 
\#54362 implies [N/Fe]$<1.55$, which is distinctly lower than the value for the most 
N-rich bright red giant observed by Yong et al. (2009). In order for this upper limit to 
coincide with such a high N abundance, our temperature scale for RHB should be lowered 
by 200~K, which we deem quite unlikely.

Oxygen: Oxygen abundances have fairly large errors, especially for BHB stars. This is because,
due to the geocentric radial velocity of NGC~1851 stars, the telluric emission
line at 7780.4~\AA\ (Hanuschik 2003) falls very close to the strongest line of the triplet.
This makes the sky subtraction uncertain for the faintest stars (the BHB ones) in our 
sample. As in the case of NGC~2808, we applied the non-LTE corrections by Gratton
et al. (1999) for RHB and RGB  stars and the RR Lyrae variable, and from Takeda 
(1997) for BHB stars.

Sodium: In all spectra the D1 line is blended with the interstellar D2 line, and cannot be 
measured accurately\footnote{There are large star-to-star variations in the strength
of the interstellar D1 line. Large EWs were obtained for stars located along two
filaments, running approximately E-W and located north and south respectively of the
cluster center. These large variations make it impossible to use e.g. the warmest HB
stars to subtract the interstellar D2 component from the blend with the stellar D1 line.}. 
In addition, we could use the doublet at 8183-94~\AA\ in all RHB and RGB stars, and 
in cooler BHB ones.

For RHB stars we adopted the non-LTE corrections by Gratton et al. (1999). For the
stars of interest here they are typically negative and small in absolute value:
$\sim -0.15$ and $\sim-0.05$~dex for the D2 line and the 8183-94~\AA\ doublet respectively.
Had we applied the updated corrections by Lind et al. (2011), these would
have been much larger in absolute value ($\sim -0.55$\ and $\sim -0.40$~dex, respectively).
However, we prefer to keep the older values by Gratton et al. (1999) for uniformity
with the analysis of red giants by Carretta et al. (2009) and of the stars in NGC~2808
(Gratton et al. 2011a). This comparison shows that quite large offsets can be
present in our Na abundances, although the star-to-star values are almost unaffected
by this uncertainty.

For the BHB stars we used the non-LTE corrections by Mashonkina et al. (2000), as
done for the BHB stars in NGC~2808. A discussion of the impact of these non-LTE
corrections can be found in Gratton et al. (2011a).

On average, the D2 lines give higher Na abundances than the doublet at 8183-94~\AA:
RHB: [Na/Fe]$_{\rm D2}$ - [Na/Fe]$_{\rm IR}=0.13\pm 0.03$~dex (30 stars, r.m.s.=0.16 dex); 
RGB: [Na/Fe]$_{\rm D2}$ - [Na/Fe]$_{\rm IR}=-0.35$~dex (1 star); 
BHB: [Na/Fe]$_{\rm D2}$ - [Na/Fe]$_{\rm IR}=0.45\pm 0.15$~dex (3 stars, r.m.s.=0.27 dex).
This difference may be attributed to the not perfect LTE corrections.

Magnesium: Mg abundances for RHB and RGB stars are based on the Mg~I line at 8213.04 
(Gratton et al. 2011a) and those for BHB on the Mg~II lines at 7877.06 and 
7896.38~\AA. For RHB and RGB stars we obtain a very small star-to-star 
scatter (only 0.06~dex), fully consistent with a single Mg abundance of 
[Mg/Fe]=$0.39\pm 0.06$. The abundance scatter is much larger for BHB stars 
(0.23 dex). This may be interpreted either as a real result
(a Mg-Na anticorrelation) or as analysis scatter.

Aluminium: The high excitation doublet at 7835.3-36.1~\AA\ was not detected in our spectra
for individual stars, either on the RGB or RHB. This sets an upper limit of [Al/Fe]$<0.2$. 
The Al feature is not even unambiguously detected in the sum of all the RHB spectra: 
in this case we detected a very weak feature which might possibly be identified with 
the strongest component of the doublet (the 7836.1~\AA\ line) with EW=1.7~m\AA, which
yields [Al/Fe]$\sim -0.6$.

Calcium: We derived LTE Ca abundances for RHB and RGB stars from the 5857 and 6122~\AA\ lines. 
Accurate oscillator strengths are available for these lines from VALD; we adopted the
same damping constants used by  Mashonkina et al. (2007). Non-LTE corrections 
are expected to be small (see Mashonkina et al. 2007). Since the lines
are quite strong, they are quite sensitive to the adopted value for the microturbulent
velocity. As mentioned in Sect 3.1, our Ca abundances for RHB stars are quite high.

Barium: The Ba abundances are based on the Ba~II line at 5853.69~\AA, which is quite strong
in the spectra of RHB and RGB stars (this line is not expected to be detectable in BHB 
stars). The line parameters adopted in our analysis are the same of Mashonkina \& Zhao 
(2006), including the collisional damping constant. Note that 
hyperfine structure should be negligible for this line (total width $<8~$m\AA).
We are not aware of statistical equilibrium computations for Ba appropriate
for this line and for model atmosphere parameters in the range of the program stars.
However, departures from LTE are not expected to be very large (see Korotin et al. 2011). 

\begin{center}
\begin{figure}
\includegraphics[width=8.8cm]{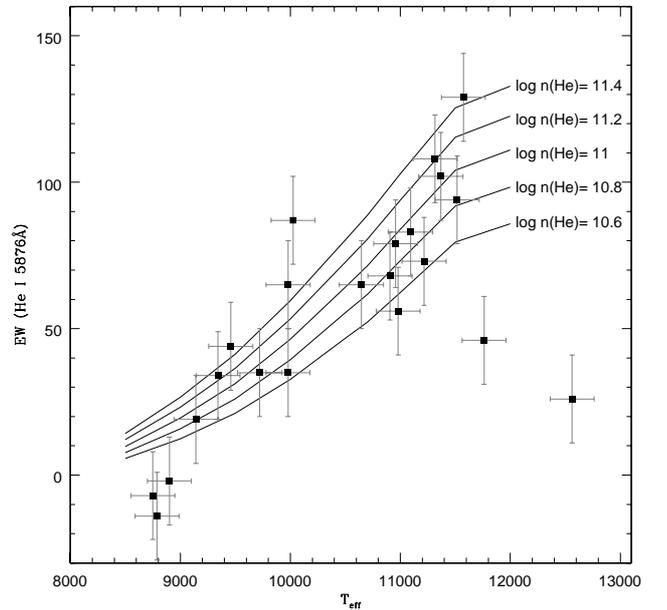}
\caption{Run of the EW of the He I line at 5876~\AA\ as a function
of $T_{\rm eff}$\ along the BHB of NGC1851. Overimposed are lines of
constant He abundance ($\log{n({\rm He})}=$10.6, 10.8, 11.0, 11.2 and 11.4).
These curves have been computed for gravities appropriate to the location
along the BHB.}
\label{f:fighe}
\end{figure}
\end{center}

\subsection{He abundances}

He abundances were obtained using the narrow multiplet at 5875.6~\AA. Figure~\ref{f:fighe} compares
the run of the EWs for this line with that expected for model atmospheres
of different temperatures, computed with parameters appropriate for the HB
of NGC~1851, and different values of the He abundances. The points relative
to individual stars display a rather large scatter, mainly due to errors 
in the EWs, which are quite large given the fairly
low S/N of the spectra. This rather large scatter precludes the use of the
He abundances to discuss properties of individual stars. However, the rather
large number of stars available allows to derive a sensible average value
of $\log{n({\rm He})}=11.01\pm 0.10$, which corresponds to an abundance in
mass of Y=$0.291\pm 0.055$. While the error bar of this average value is still 
quite large, it agrees fairly well with expectations for a population with an 
initial He abundance as given by the Big Bang alone (Y=0.248: Cyburt 2004) and 
later modified by the effect of the first dredge-up, which is $\Delta Y\sim 0.015$\ 
for the stars under consideration (Sweigart 1987). The error bar is large enough 
to accommodate a moderate He enhancement; however very large initial He abundances 
($Y>0.33$) are not compatible with the present result. This agrees with the lack 
of evidence for a broadening of the main sequence (Milone et al. 2008). A normal
helium was also found using the R-parameter (Salaris et al. 2004).

\subsection{Error analysis}

\begin{table*}[htb]
\centering
\caption[]{Sensitivity and error analysis. Variation is the change in each parameter
used in the sensitivity analysis (Column 2-6), while error is the change used
to estimate the total error (Column 7).}
\begin{tabular}{lcccccc}
\hline
Parameter &$T_{\rm eff}$&$\log{g}$&$[$A/H$]$&$v_t$& EW & Total \\
\hline
Variation& +100 K & +0.5 dex & +0.1 dex  & +0.5 km/s &+10 m\AA &     \\
Error BHB& 200 K & ~0.1  dex & ~0.05 dex & ~0.5 km/s &~13 m\AA &        \\
Error RHB& ~50 K & ~0.05 dex & ~0.05 dex & ~0.5 km/s &~~8 m\AA &        \\
Error RGB& ~50 K & ~0.05 dex & ~0.05 dex & ~0.5 km/s &~~8 m\AA &        \\
\hline
\multicolumn{7}{c}{Blue HB}\\
\hline
$[$N/Fe$] $ &  0.018 & ~0.049 & -0.001 & -0.057 & 0.092 & 0.138 \\ 
$[$O/Fe$] $ &  0.021 & -0.002 & -0.012 & -0.154 & 0.099 & 0.205 \\ 
$[$Na/Fe$]$ &  0.087 & -0.169 &  0.000 & -0.057 & 0.162 & 0.281 \\ 
$[$Mg/Fe$]$ &  0.000 & ~0.052 & -0.004 & -0.056 & 0.144 & 0.196 \\ 
\hline
\multicolumn{7}{c}{Red HB}\\
\hline
$[$Fe/H$]$  & ~0.065 & -0.020 & -0.001 & -0.098 & 0.050 & 0.078 \\ 
$[$O/Fe$]$  & -0.095 & -0.181 & ~0.000 & -0.081 & 0.080 & 0.095 \\ 
$[$Na/Fe$]$ & ~0.095 & -0.125 & ~0.003 & -0.126 & 0.049 & 0.098 \\ 
$[$Mg/Fe$]$ & ~0.025 & -0.015 & ~0.000 & -0.022 & 0.160 & 0.129 \\ 
$[$Si/Fe$]$ & ~0.030 & ~0.005 & ~0.001 & -0.027 & 0.094 & 0.078 \\ 
$[$Ca/Fe$]$ & ~0.075 & -0.100 & ~0.000 & -0.200 & 0.158 & 0.155 \\ 
$[$Ba/Fe$]$ & ~0.056 & ~0.181 & ~0.006 & -0.169 & 0.181 & 0.180 \\ 
\hline
\end{tabular}
\label{t:tab4}
\end{table*}

Error analysis was done in the usual way, by repeating the abundance
derivation by modifying a single parameter each time. Relevant data
are given in Table~\ref{t:tab4}. The last column gives an estimate of
the total internal errors obtained using the sensitivities listed above,
as well as the errors in the individual parameters given on lines 2 and
3 for blue and red HB stars, respectively. 

\begin{center}
\begin{figure*}
\includegraphics[width=18.0cm]{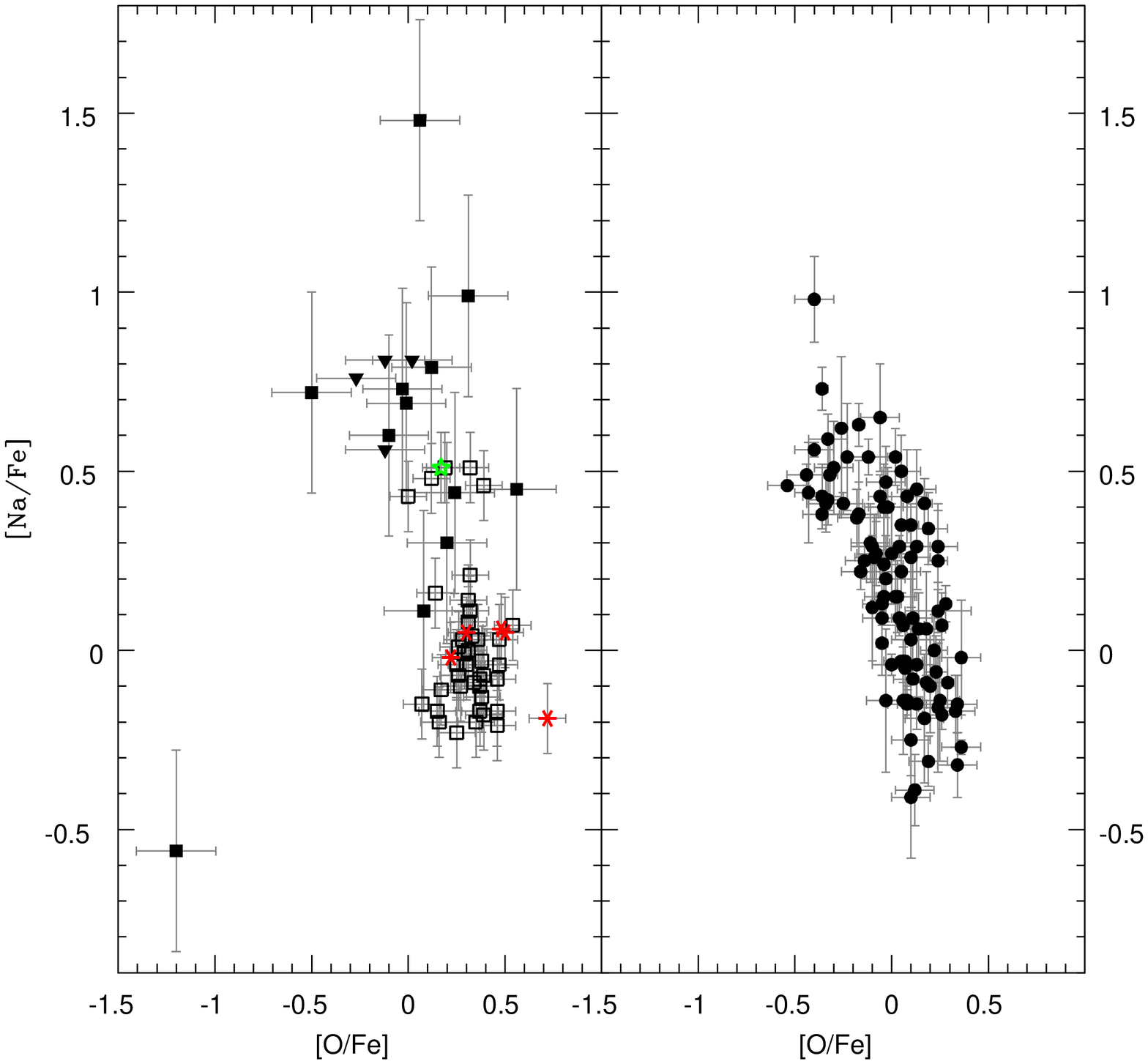}
\caption{Left panel: Na-O anticorrelation for HB stars of NGC~1851. Filled squares are
BHB stars (filled triangles are upper limits for Na); the open star is the RR Lyrae variable;
open squares are RHB stars; the asterisks are red giants. Right panel: the same distribution
for RGB stars from Carretta et al. (2011b).}
\label{f:fig3}
\end{figure*}
\end{center}

\begin{center}
\begin{figure}
\includegraphics[width=8.8cm]{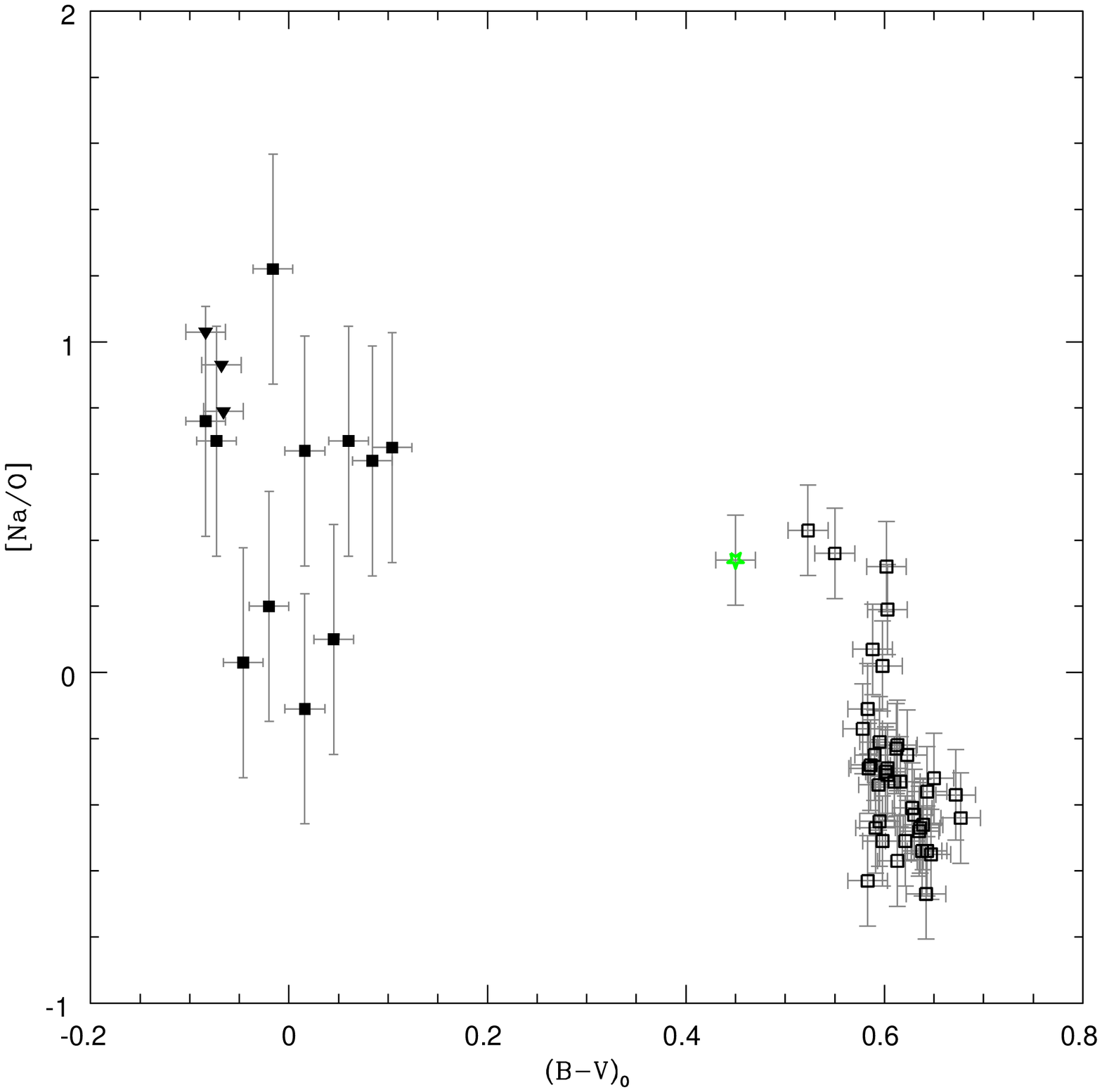}
\includegraphics[width=8.8cm]{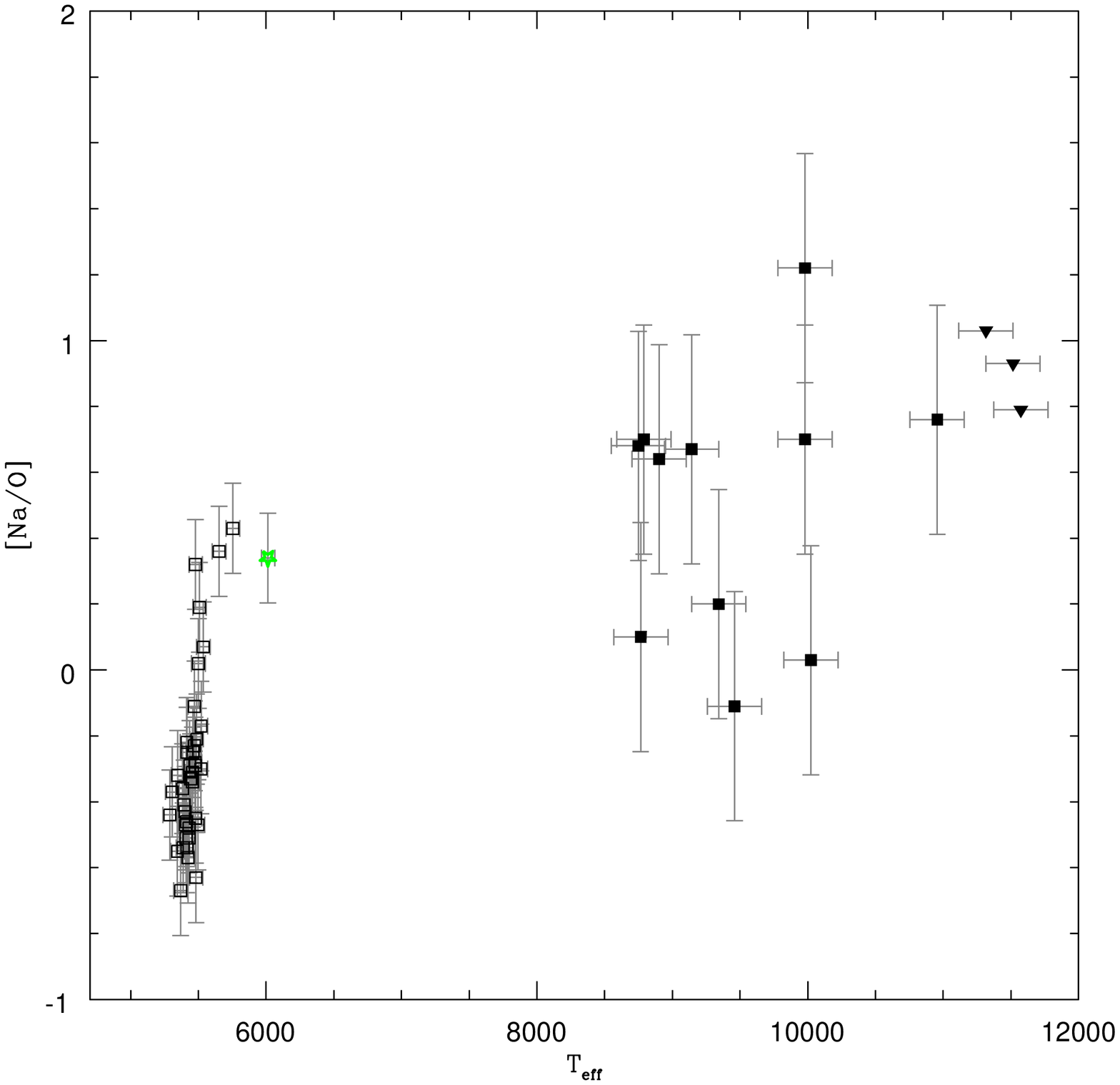}
\caption{Upper panel: Run of the [Na/O] abundance ratio with $B-V$\ colour along the
HB of NGC~1851 (left panel). Lower panel: the same, but with $T_{\rm eff}$.
Symbols are the same as in Fig.~\ref{f:fig3}.}
\label{f:fig4}
\end{figure}
\end{center}

\begin{center}
\begin{figure}
\includegraphics[width=8.8cm]{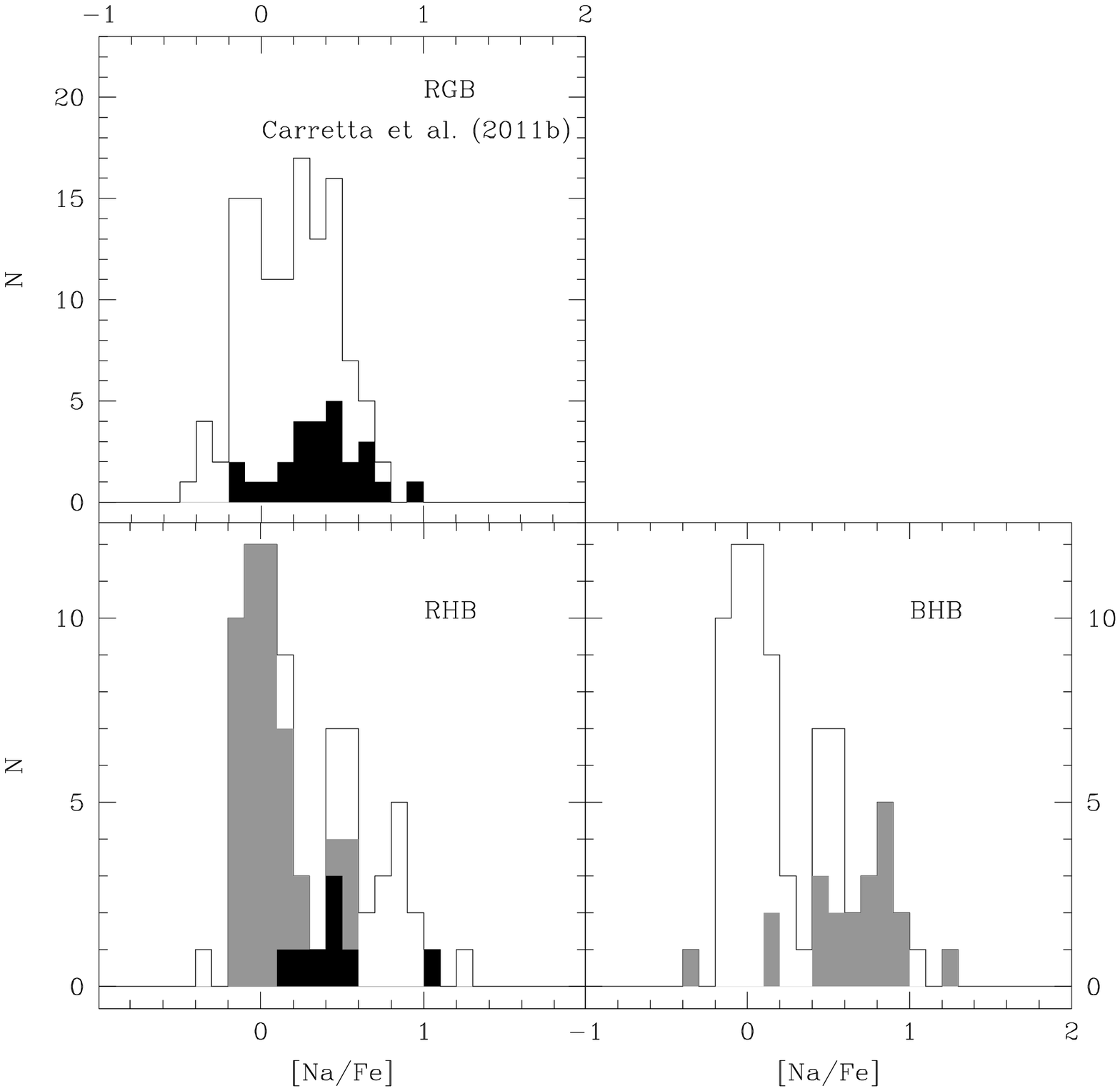}
\caption{Distribution of [Na/Fe] abundances for the HB stars analyzed in this 
paper (bottom panels) and for the RGB stars of Carretta et al. (2011b) (upper 
panel). Grey histograms indicate the distribution of RHB (bottom-left panel) 
and BHB (bottom-right panel) stars. Black histograms indicate the distribution 
of stars with [Ba/Fe]$>0.6$.}
\label{f:nadist}
\end{figure}
\end{center}

\section{Discussion}

\subsection{The Na-O anticorrelation}

Figure~\ref{f:fig3} shows the Na-O anticorrelation we obtain for the HB stars 
of NGC~1851. Different symbols are used for blue and red HB stars. For
comparison, we also plotted the Na-O anticorrelation for red giants by 
Carretta et al. (2011). We remind that the HB stars hotter than 11,500~K are 
not considered here because their surface abundances are not related in a 
simple way to their original composition. However, only $\sim 5$\% of the HB
stars of NGC~1851 are that warm.

On the whole, we obtain a clean Na-O anticorrelation with an inter-quartile 
of IQR(Na/O)=$0.70\pm 0.41$, which is very similar to the value obtained 
for the RGB by Carretta et al. (2011: IQR(Na/O)=0.69). On the other hand, the 
relation between O and Na abundances with colour/temperature is not as 
clear as it was the case for NGC~2808 and there is considerable overlap 
between the range of O and Na abundances covered by BHB and RHB stars (see
Figure~\ref{f:fig4}).

More in detail, we compare in Fig~\ref{f:nadist} the distribution in [Na/Fe] 
of BHB and RHB stars. For comparison the same distribution using the abundances of 
RGB stars by Carretta et al. (2011b) are also shown. It is apparent that RHB 
stars distribute in two well defined groups: a "majority"
group (including 36 stars) have [Na/Fe]$<0.3$\ and a small group (5 stars)
have [Na/Fe]$>0.3$. The statistics is improved by considering also the
stars that have not been observed with HR19, and then have no O abundance.
The subdivision in two groups is still very clean after this addition: there are
in total 46 Na-poor and 8 Na-rich stars. In addition, one star (\#31903) has a 
very high [Na/Fe]=0.93. 

The RR Lyrae variable has Na and O abundances very close to that found for the
group of Na-rich and moderately O-poor RHB stars.

The BHB stars also define a Na-O anticorrelation, including both O-rich 
([O/Fe]$\sim 0.3$) stars with moderate Na excesses ([Na/Fe]$\sim 0.5$) and a 
prominent group of O-poor ([O/Fe]$\sim 0.2$) and Na-rich ([Na/Fe]$\sim 0.7$) stars. 
However, the large errors associated to individual BHB stars do not allow to 
understand if the distribution is continuous or bimodal: this ambiguity is even
more evident if we consider also the stars for which no information about the
O abundance was available, which spread over a large range of Na abundances
(see Fig.~\ref{f:nadist}).
There are even two stars (\#28078 and \#40227) for which we get [Na/Fe]$>1$.

On the whole, we obtain a correlation between the [Na/O] abundance ratio
and colour of the stars (see Figure~\ref{f:fig4}). However, while this correlation is very clean for the
RHB stars (the Spearman ranking test gives a probability smaller than 0.05\%
that the observed correlation is random), it is much less clear within the BHB ones
(in this case the probability is only smaller than 7\%), although on average they
have lower O and higher Na abundances (for the whole sample,
the probability of a random result is smaller 
than 0.05\%; see Figure~\ref{f:fig4}).
To further complicate the interpretation of observational data, we must take
into account the possibility of systematic offsets between the abundances
obtained for RHB and BHB stars, mainly related to uncertainties in the
non-LTE correction. This result of our analysis should then be taken with some
caution, and it is well possible that there is no real offset in Na abundances
between RHB and BHB stars. However, we deem the conclusion that 
O-Na anticorrelations do exist separately for both BHB and RHB stars robust.

\subsection{CNO in BHB stars}

Cassisi et al. (2008) and Ventura et al. (2009) suggested that the f-SGB is due to a CNO rich
population that is also responsible for the BHB. We may compare the
prediction of this hypothesis (the BHB are CNO-rich) with our results.
We obtain average values of [N/Fe]=+1.16 and [O/Fe]=0.00 for the BHB stars.
We have not determined the C abundances. However, both Yong et al. (2009)
and Villanova et al. (2010) have obtained C abundances for several stars
on the RGB, and did not find any stars with [C/Fe]$>-0.2$; the average
values are [C/Fe]$\sim -0.7$\ in both studies, with little dispersion.
If we then assume that C gives a negligible contribution to
the total CNO abundance in the BHB stars of NGC 1851, we find a total of 
[(C+N+O)/Fe]$\sim 0.24$, a value which is fairly typical for metal-poor stars. 
Since we get a lower limit of [(C+N+O)/Fe]$>0.12$\ simply considering the O 
abundance of the Na-poor stars, we conclude that there is no evidence 
for a significant ($\Delta$[(C+N+O)/Fe]$>0.2$)\ excess of CNO elements among BHB stars in NGC~1851, at variance
with the prediction of the Cassisi et al. and Ventura et al. scenario.

\begin{center}
\begin{figure}
\includegraphics[width=8.8cm]{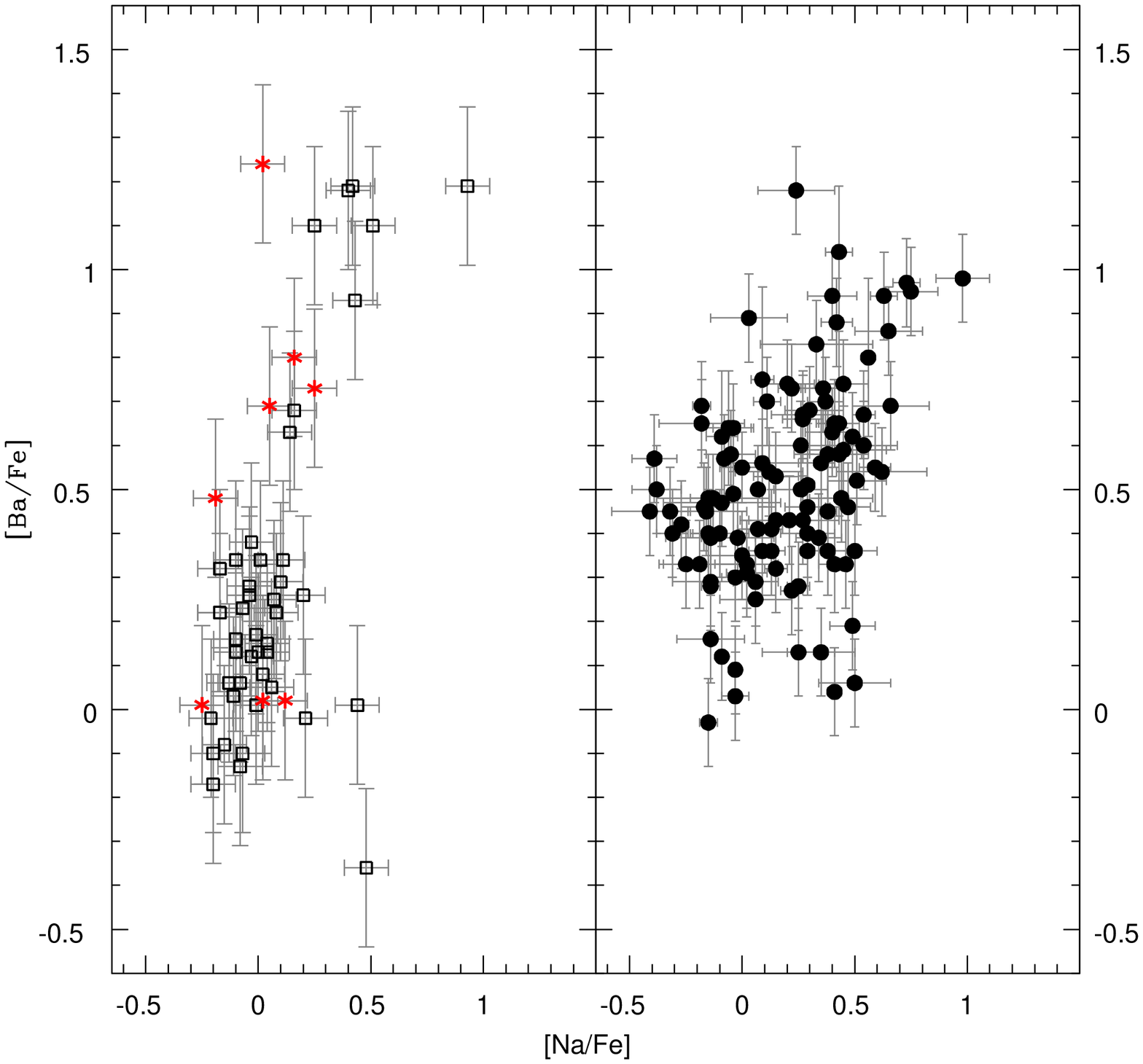}
\caption{Correlation of the [Na/Fe] and [Ba/Fe] abundance ratios for stars
of NGC~1851. Left panel: HB stars; right panel: RGB stars from Carretta
et al. (2011b). Symbols are the same as in Fig.~\ref{f:fig3}.}
\label{f:bana}
\end{figure}
\end{center}

\subsection{Ba and N abundances in RHB stars and their relation with Na abundances}

The Ba abundances also provide useful information in NGC~1851. Yong et al. 
(2008, 2009), Villanova et al. (2010), and Carretta et al. (2011a) have 
obtained clear correlations between Na and Ba abundances and have shown that
the large Ba abundances found in several RGB stars of NGC~1851 can be
attributed to the $s-$process. This suggests an important contribution
by thermal pulsing AGB stars to the chemical evolution of this GC. Here, we 
measured Ba abundances for 44 RHB and eight RGB stars but we have no information 
about Ba in the BHB stars and the RR Lyrae variable. We obtain a clear 
correlation between Na and Ba abundances (see Figure~\ref{f:bana}). A small 
part of this correlation might be explained by scatter in the appropriate 
value for the microturbulent velocity for individual stars, since these 
lines are quite strong and we adopted the same value for this parameter 
in all stars. However, the correlation extends over a very broad range 
and extreme values cannot be explained in this way. 

We recall that Carretta et al. (2011a) found a large spread in Ba abundances 
along the RGB. Ten stars in their sample ($9\pm 3$\%) are clearly Ba-rich 
([Ba/Fe]$>0.7$), while the remaining ones distribute over the range 
0.1$<$[Ba/Fe]$<$0.7. This latter scatter might be related to the use of a not
optimal line (the 6141~\AA\ one, which is blended with a Fe~I line). However, 
in that work the group of Ba-rich stars tends to have on average larger
abundances of Na (0.2$<$[Na/Fe]$<0.5$; see Fig. \ref{f:nadist}). On the other 
hand, our RHB stars clearly divide into two groups: 
seven stars are Ba-rich ([Ba/Fe]$>0.7$), the remaining are Ba-poor 
([Ba/Fe]$<0.4$). Since RHB makes up $\sim $60\% of the HB stars in NGC~1851, 
Ba-rich RHB stars should be $\sim 10\pm 4$\% of the stars, while Ba-poor 
correspond to $\sim 50\pm 4$\%. The identification of Ba-rich RHB and RGB 
stars is quite obvious. For exclusion, we then suggest that the BHB stars 
(for which we did not measure Ba abundances)
join the remaining RHB stars being Ba-poor.

We notice that the fraction of Na/Ba-rich RGB stars 
is consistent with the fraction of stars 
along the anomalous $v-y$\ sequence ($\sim 7$\%: see Carretta et al. 2011b).
Indeed all stars on the anomalous red RGB in the $v,~(v-y)$ diagram are 
Ba-rich (Villanova et al. 2010; Carretta et al. 2011a). The Str\"omgren 
colours of this sequence indicate very strong CN bands, which might be 
explained either by an extremely large N abundance
\footnote{We estimated the impact of a very large N abundance on $v-y$\
computing synthetic Str\"omgren colours for for a red giant with the
same approach of Carretta et al. (2011b). We considered the case of a red
giant with $M_V=-1$, [Fe/H]=-1.3, [C/Fe]=-0.6, [N/Fe]=1.7, and [O/Fe]=-0.1,
and compared its $v-y$ colour with those obtained for a more normal
composition ([C/Fe]=-0.6, [N/Fe]=1.4, and [O/Fe]=-0.1). We found that an
extremely N-rich star has a $v-y$ colour 0.41 mag redder than the "normal
N-rich" one. This comparison shows that the anomalous $v-y$\ sequence
can be produced by [N/Fe]$\sim 1.55$, which implies a CNO excess of only
$\sim 0.15$~dex with respect to normal stars in NGC~1851.}
, or by a C abundance 
comparable to that of O in the atmospheres of these stars, leaving a lot 
of C available for the formation of C-bearing molecules such as CN and CH 
in cool RGB stars (see Carretta et al. 2011b). In the second case, since 
Ba-rich stars are Na-rich and moderately O-poor, C abundances do not need 
to be extraordinary large, and even a moderate excess of C+N+O might explain 
observations. The direct determination of the CNO content for a few stars of 
this sequence by Yong et al. (2011) indicates a low C ([C/Fe]$\sim -0.7$) and a 
very high N content (up to [N/Fe]$\sim +2.2$), favouring the first explanation,
since they find [(C+N+O)/Fe]$\geq 1.0$\ for the most extreme stars. The same 
Yong et al. (2011) caution however that use of NH rather than CN (the 
specie considered for the NGC~1851 analysis) provide fairly lower N 
abundances (up to 0.44~dex) for RGB stars in NGC~6752. Were this same difference
be valid for NGC~1851, a less spectacular but still quite high [(C+N+O)/Fe]$\sim 0.6$\ 
would be obtained for the same stars. Villanova et al. (2009) only published the 
sum of the CNO abundances 
(which they found to be [(C+N+O)/Fe]$\sim 0.2$\ for both the Ba-rich and Ba-poor 
stars), but they kindly provided us with the values they obtained for the 
individual elements. They also found low C abundances in all the stars they 
examined, the mean value for the stars on the Ba-rich sequence being [C/Fe]=-0.58. 
For the same stars they also found [N/Fe]=1.09 and [O/Fe]=-0.15. For stars in the 
Ba-poor sequence they obtain [C/Fe]=-0.88, [N/Fe]=0.71 and [O/Fe]=0.09. While C 
and O abundances agree fairly well with those of Yong et al., the N abundances 
are very different, which is surprising since the same CN lines 
are used in both analyses. We conclude that the exact values of the CNO abundances 
in the RGB Ba-rich stars is still uncertain, although they are definitely more 
N-rich and C- and O-poor than the RGB Ba-poor stars.

We have not measured the sum of C+N+O for the RHB stars. However, the lack of
detection of the N~I line at 8216~\AA\ yields an upper limit to the N abundance 
of [N/Fe]$<1.55$\ for the warm Ba-rich star \#54362, which has [O/Fe]=0.00. 
Given the possible systematic errors in N abundances from CN lines and the
error bar of our determination, this is perhaps not incompatible with the 
results of Yong et al. for Ba-rich RGB stars, and agrees with that of Villanova 
et al. If we now assume that the C contribution to the sum of C+N+O is negligible, 
as found for stars along the RGB, we get an upper limit of 
[(C+N+O)/Fe]$<0.5$\ for this star. Taken literally, the comparison between the upper 
limit for the Ba-rich RHB star and the BHB ones indicates an excess of C+N+O 
smaller than a factor of 2 for the first star. This is lower than the range in
CNO abundances estimated by Yong et al., and in agreement with that by Villanova 
et al. On the other hand, assuming [(C+N+O)/Fe]$\sim 0.15$\ is however enough
to explain the anomalous $v-y$ colours of these stars.
There are several possible sources of errors in this determination, including 
systematic errors in the atmospheric parameters or departures of real atmospheres 
from the model ones. Also, it is possible that star \#54362 is not a typical 
Ba-rich star (the Ba excess of [Ba/Fe]=+0.93 is lower than that obtained for a 
few other stars). However, this comparison suggests that after all the Ba-rich 
RHB stars might possibly be not particularly rich in the sum of C+N+O.

\subsection{The Na/Ba poor and Na/Ba-rich RHB stars in the colour-magnitude diagram} 
 
The Na-poor RHB stars have a very small range in $B-V$\ and Str\"omgren colours, 
and hence $T_{\rm eff}$\ ($\sim 200$~K peak-to-valley). They appear as a very 
compact group in all diagrams we plotted. The Ba-rich RHB stars are bluer
(by $0.044\pm 0.012$~mag in $B-V$), warmer (by $148\pm 32$~K), and on average 
slightly brighter than the other RHB stars: the difference in $V$\ magnitude is 
small ($-0.035\pm 0.012$~mag) but significant at almost 3$-\sigma$. The two 
groups of RHB stars separate clearly in the colour-magnitude diagram, because 
differences of average values are larger than the internal scatter of each group. 

The luminosity difference between Ba-rich and Ba-poor RHB stars might in
principle be explained in various ways. For instance, it might be
attributed to a difference of $\Delta Y\sim 0.008$\ in the He content. 
However, a difference in He abundance alone would not explain why the RGB 
Ba-rich stars (likely the progenitors of the RHB Ba-rich stars) have anomalous 
$v-y$\ colours. This hypothesis is then not enough to justify all observations. 

A difference in HB luminosity similar to that observed
would also be produced by a change of 0.1~dex in metallicity, were
the Na-rich more metal-poor than the Na-poor ones. However
Na-rich RHB stars have [Fe/H]=$-1.121\pm 0.015$, r.m.s=0.045 dex,
and Na-poor ones [Fe/H]=$-1.147\pm 0.010$, r.m.s=0.066 dex.
This is not likely to produce appreciable differences in both
luminosity and colours. Therefore this hypothesis does not agree 
with observations. 

Furthermore, since CNO abundance variations have been
widely proposed to explain the SGB and HB of NGC~1851 (Salaris et al. 2008; 
Cassisi et al. 2008; Ventura et al. 2009), it is useful to consider if they can justify these
observations. However, while CNO-rich HBs are indeed brighter than CNO-normal
ones, they are expected to be also much redder (Lee et al. 1994; Pietrinferni 
et al. 2009), while they are bluer. This solution is then not
acceptable too.

We finally note that there are two Na-rich and Ba-poor RHB stars 
(\#39317 and \#50923). They are $\sim 300$~K warmer and $\sim 0.1$~mag
brighter than the average RHB stars. We suggest that these two
stars evolved well off their ZAHB locations, which were possibly
on the BHB or within the instability strip. Incidentally, the
only RR Lyrae variable is Na-rich, but unluckily we do not have
Ba abundance determination for this star.
 
\begin{center}
\begin{figure*}
\includegraphics[width=18cm]{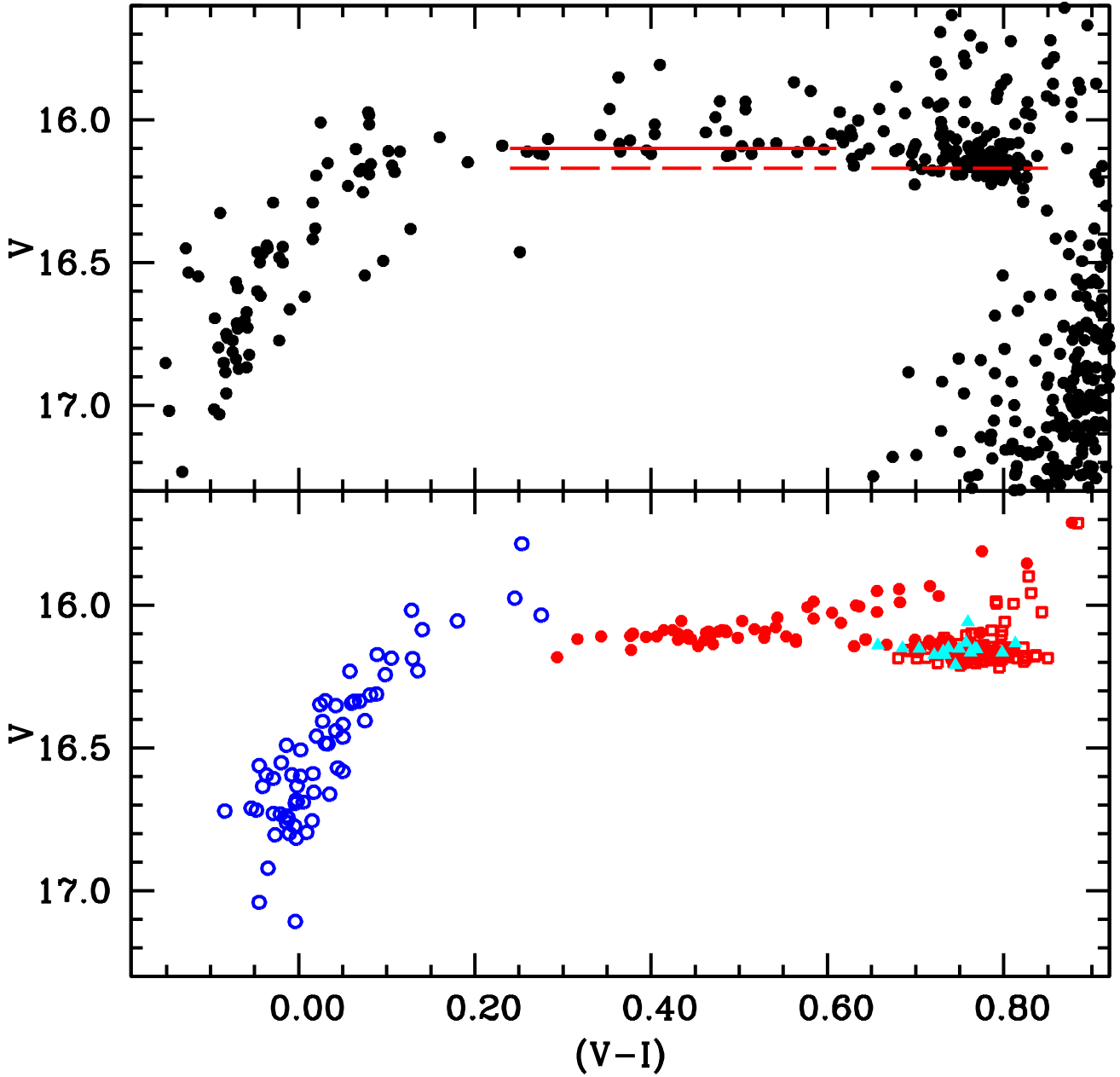}
\caption{Comparison between observed (upper panel) and a synthetic HB
for NGC~1851. The synthetic CMD has been obtained by considering:
$<M>$=0.670$\pm$0.005 $M_{\odot}$, Y=0.248 (red open squares); 
$<M>$=0.640$\pm$0.006 $M_{\odot}$, Y=0.265 (filled red circles);
$<M>$=0.650$\pm$0.004, $M_{\odot}$, Y=0.248 (cyan filled triangles);
$<M>$=0.590$\pm$0.005 $M_{\odot}$, Y=0.280 (blue open circles). These
populations have been selected in order to reproduce the observed
distribution of stars on the HB and SGB, as well as the anomalous red
sequence on the RGB in the $(v, (v-y))$\ diagram (see text for further details). 
Solid and dashed lines in the upper panel 
represent the average brightness of, respectively, 
the horizontal part of the blue HB and the red HB, as derived from the synthetic CMD. For ease of 
comparison, the dashed line is also displayed at the colour range of the horizontal part of the blue HB.}
\label{f:simhb}
\end{figure*}
\end{center}

\subsection{Comparison with simulated HBs}

To look for an explanation of the conundrum of the Ba-rich RHB stars, we have 
to consider differences in more than a single parameter between Ba-rich and 
Ba-poor RHB stars. For instance, we may assume that the Ba-rich stars are 
not only more rich in CNO elements (which may explain the anomalous
$v-y$\ colour on the RGB), but also older (and then less massive) than 
the Ba-poor RHB stars. In the scenario where the separation between BHB/f-SGB 
and RHB/b-SGB sequences is due to an age difference of about 1 Gyr,
this corresponds to attributing to the Ba-rich stars an age more similar 
to the first group or at least intermediate between the two. To explore 
this possibility, we ran some simulation for the HB of NGC~1851.
They were performed as described in Salaris et
al.~(2008) and we adopted Walker~(1998) $VI$\ photometry to create the  
reference observational CMD, for it includes also mean magnitudes of RR Lyrae stars, that 
fill the gap seen in Fig.~\ref{f:fig2} between the RHB and the more extreme BHB. We will denote 
as 'horizontal part of the BHB' this section of the observed HB. 

We have employed as reference set the HB evolutionary tracks
for [Fe/H]=$-$1.31, Y=0.248, [$\alpha$/Fe]=0.4 from the BaSTI database\footnote{http://www.oa-teramo.inaf.it/BASTI}
(Pietrinferni et al.~2006). In addition, we have interpolated among the
$\alpha$-enhanced BaSTI models at Y=0.248 and Y=0.300, to determine HB
tracks for intermediate values of Y, at [Fe/H]=$-$1.31.
Finally, we have also interpolated between the reference set and the
CNO and Na anticorrelated models with CNO sum enhanced by 0.3~dex
(Pietrinferni et al.~2009) to determine HB tracks with a milder
CNO-enhancement, equal to 0.15~dex. We adopted E(B-V)=0.02 (Walker~1998) 
and employed a distance modulus $(m-M)_V$=15.58 obtained from matching the 
mean magnitude of the RHB with our synthetic counterpart.

In our simulations we have considered as constraints a number
ratio between b-SGB and f-SGB stars equal to 70:30, and
a 30:10:60 ratio between stars at the blue side, within, and at the
red side of the instability strip, respectively.
Figure~\ref{f:simhb} 
displays a realization of our synthetic calculations of the cluster HB (bottom panel) 
compared to the observed CMD (upper panel). The number of stars in the 
synthetic HB is approximately equal to the observed numbers.
We considered the observed HB of NGC1851 formed by the contribution 
of four different stellar populations, with a Gaussian mass distribution for 
each component:

\begin{itemize}
\item The BHB/f-SGB population: it is associated to the f-SGB and 
makes up therefore $\sim$25\% of the cluster stellar content. 
To match both color and magnitude of the BHB we adopted  
a HB mass of $< M >=0.590\pm0.005 M_{\odot}$ (corresponding to a total 
mass loss of the RGB progenitor equal to $\sim$0.20$M_{\odot}$ for a $\sim$12 Gyr 
population with the chemical composition 
specified below) a normal [C+N+O]/Fe 
abundance and an helium content Y=0.280 (blue circles in Fig.~\ref{f:simhb});
\item The RHB population: this population is the dominant cluster 
population ($\sim$55\%) and represents the vast majority of 
the progeny of b-SGB stars. It 
has been assumed to be $\sim$ 1.5~Gyr younger than the BHB one, with a 
cosmological helium content Y=0.248 and the same [C+N+O]/Fe abundance 
of the BHB. The observed CMD is reproduced with 
a HB mass of $<M>=0.670\pm0.005 M_{\odot}$ (open red squares in the 
Fig.~\ref{f:simhb}) consistent with the same total RGB mass loss as for the 
progenitors of BHB objects;
\item The population of Ba-rich stars: this population is $\sim$10\% 
of the cluster stellar content. It has been reproduced with a 
mass of  $<M>=0.650\pm0.004 M_{\odot}$ (consistent with an intermediate 
age between the two above populations), assuming a cosmological helium content 
Y=0.248 and an enhanced [C+N+O]/Fe by 0.15 dex (cyan filled triangles 
in the Figure~\ref{f:simhb}). This mean mass corresponds to 
a progenitor age intermediate between the RHB and BHB populations, when the same total RGB mass loss 
is assumed;
\item A fourth population constituting $\sim$10\% of the cluster stars 
has been added to populate the horizontal part of the BHB 
visible in Walker~(1998) CMD, that includes the 
instability strip (filled red circles in Figure~\ref{f:simhb}). We 
employed for these stars a HB mass of $<M>=0.640\pm0.006 M_{\odot}$, 
Y=0.265 and normal [C+N+O]/Fe abundances, to reproduce both the color extension and 
the magnitude of the HB at $0.25<V-I<0.55$. 
Our spectroscopic data do not cover this portion of the HB, but this additional population is 
obviously necessary to 
reproduce the observed HB in the region of the instability strip, and 
accounts for the two Na-rich Ba-poor stars brighter and warmer than the bulk of RHB stars. 
\end{itemize}

Our synthetic HB simulations show that with a suitable choice of age and CNO abundances we can reproduce
the location of the Ba-rich stars in the color-magnitude diagram for
both RGB and HB stars. Other pairs of parameters are not as successful.
The anomalous $v-y$\ colours of the RGB Ba-rich stars can only be
explained with overabundances of CNO elements. The small differences in 
luminosity and [Fe/H] rule out pairs such as (He, CNO) or ([Fe/H], CNO).

\begin{center}
\begin{figure}
\includegraphics[width=8.8cm]{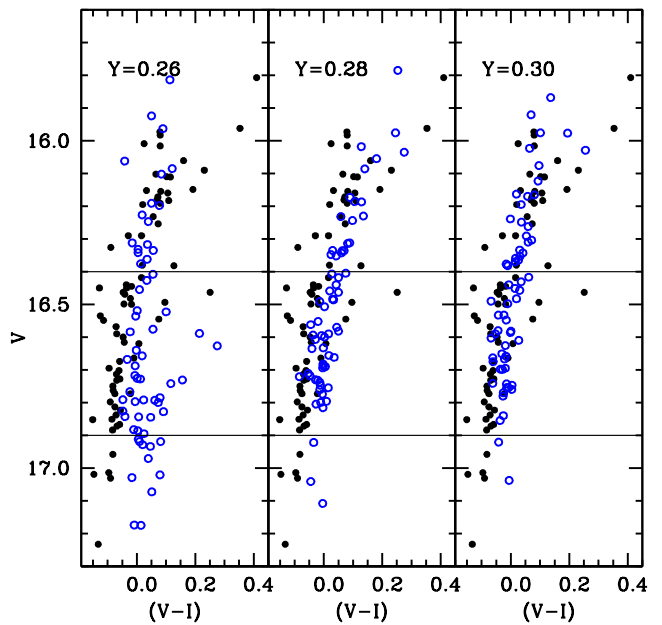}
\caption{Comparison between the photometry displayed in Fig.~\ref{f:simhb} restricted to the
bluest HB stars, and the corresponding synthetic HB simulations, for various assumptions about the initial He contents (see labels).
The two horizontal lines mark the region in the CMD that include 
stars whose surface He abundance have been measured for this work.}
\label{f:simbluehb}
\end{figure}
\end{center}

\begin{center}
\begin{figure}
\includegraphics[width=8.8cm]{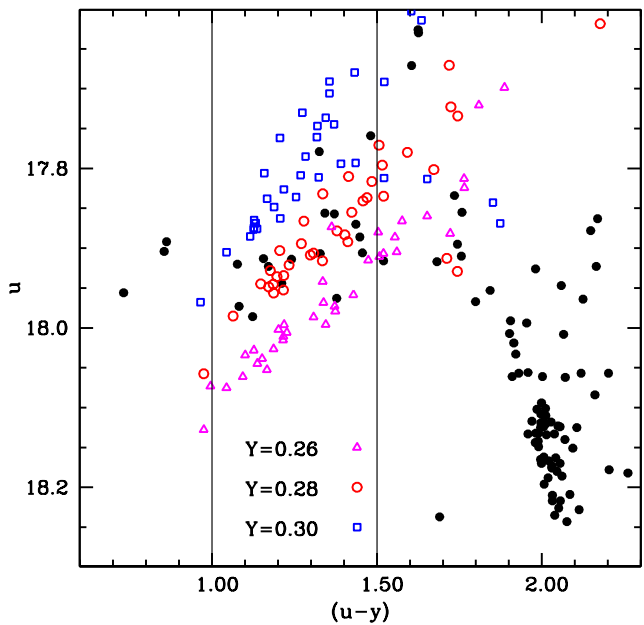}
\caption{As Fig.~\ref{f:simbluehb} but in the Stroemgren $u, (u-y)$ CMD. 
The adopted distance modulus and reddening
are the same employed in Fig.~\ref{f:simhb}. The two horizontal lines have the same meaning as in Fig.~\ref{f:simbluehb} }
\label{f:simhbstroe}
\end{figure}
\end{center}

As outlined above, our selected distribution of initial He abundances generates a satisfactory fit 
to the overall HB morphology.
In particular, the initial He adopted for the bulk of RHB stars and for the horizontal part of the blue HB 
reproduces nicely the increase of the average HB brightness when moving away from the RHB towards 
higher effective temperatures. 
On the other hand, as shown by Fig.~\ref{f:simbluehb},  
the bluest HB is almost perpendicular in the $V, (V-I)$ plane, and does not allow a clear-cut 
selection of the most appropriate
initial He abundance. One can safely conclude that the lower He mass fraction Y=0.26 is clearly ruled out, 
and there is some hint that Y=0.30 generates a too steep sequence compared to the observed CMD, but some 
constraints are required. 
They are indeed provided by CMDs that make use of  
ultraviolet photometric bands, as has been shown conclusively by Busso et al.~(2007) and Dalessandro et al.~(2011). 
We have therefore compared our 
synthetic HB simulations with data in the Str\"omgren $u, (u-y)$ CMD -- obtained from the same dataset used for 
determining the $T_{eff}$ of our star sample (see the discussion in the previous sections). The comparison is shown in 
Fig.~\ref{f:simhbstroe}. 
In this CMD the morphology of the bluer part of NGC1851 HB puts strong constraints on the initial Y values:  
both the lower (Y=0.26) and higher (Y=0.30) He abundances are clearly ruled out, whereas an initial He abundance Y=0.28 
is able to reproduce the observed distribution of stars in this CMD. This value of Y is consistent 
with the central value of the He abundance distribution obtained from spectroscopy.
Notice the three stars located at $(u-y) < $1.0,  
that appear overluminous compared to trend set by the synthetic calculations. Their colors correspond to 
$T_{eff}$ above 12000~K, that marks the onset of radiative levitation (Grundahl et al.~1999).
It is also important to remark that at the relatively high temperatures of the BHB stars displayed in 
Fig.~\ref{f:simhbstroe} the effect of the CNONa abundance anticorrelations on the bolometric corrections 
to the $u$ and $y$ bands are expected to be negligible (see Sbordone et al.~2011).

This discussion shows that many aspects of NGC~1851 can be derived by using the whole set of
observations available for HB stars. However, we acknowledge that we had to consider many
different stellar populations, and this may then appear quite contrived. In addition, we are 
aware that other combinations of parameters are likely possible, including
e.g. suitable mass loss laws for different group of stars. Anyhow, we think that any model
trying to reproduce this whole set of observations must assume that NGC~1851 contains
many stellar populations and that its history was certainly complex. 
Finally, we wish to notice that the difference of our derived HB He abundance distribution
compared to Salaris et al.~(2008) results 
is mainly due to the different photometric datasets we employed here, 
as well as to the larger number of observational constraints
accounted for in our simulations.

\begin{center}
\begin{figure}
\includegraphics[width=8.8cm]{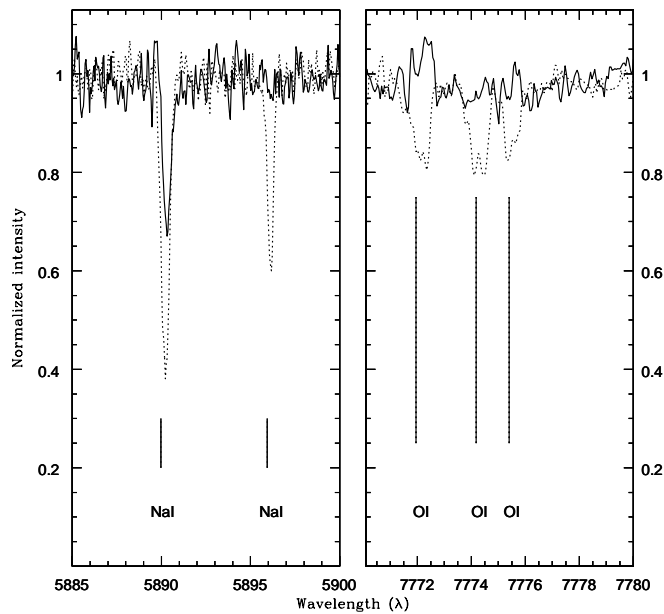}
\caption{Comparison between the spectrum of star \#46902 (solid line) and that
of a star with very similar temperature (\#48007, dashed line) in the regions of
the Na~D (left panel) and O~I lines (right panel). Note that the 
stellar D1 line at 5890~\AA\ is blended with the interstellar D2 line
in these spectra. }
\label{f:fig46902}
\end{figure}
\end{center}

\begin{center}
\begin{figure}
\includegraphics[width=8.8cm]{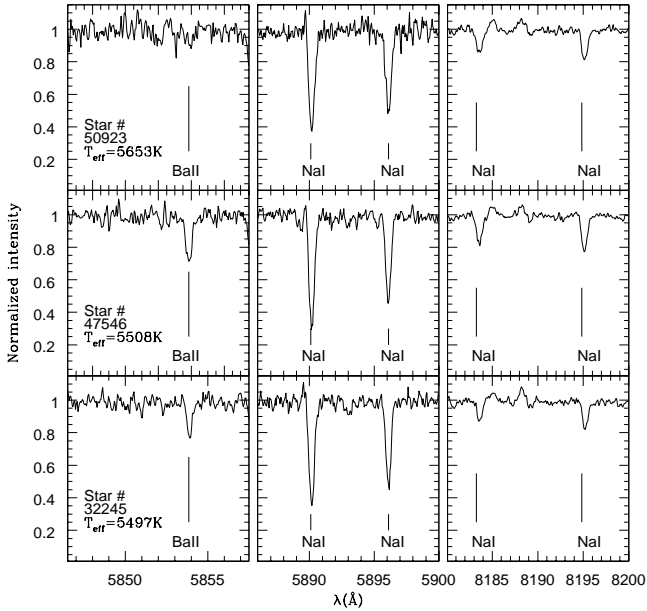}
\caption{Comparison between the spectra of three RHB stars:
star \#50923 is a Na-rich and Ba-poor star; star \#47546 is 
Na-rich and Ba-rich; star \#32245 is Na-poor and Ba-poor. 
Note that the stellar D1 line at 5890~\AA\ is blended with the 
interstellar D2 line in these spectra. }
\label{f:figspectra}
\end{figure}
\end{center}

\subsection{Peculiar stars}

It is very unlikely that stars with radial velocities compatible with that 
of NGC~1851 are not member of the cluster. For this reason, stars with 
peculiar abundances should be examined carefully. There are a few such stars.
Along the BHB, star \#46902 has very low abundances of both O and Na
([O/Fe]=-1.20, [Na/Fe]=-0.56). Other stars of NGC~1851 with a similar 
temperature have much stronger lines (see Figure~\ref{f:fig46902}); we notice that the 
Na D2 line is however detectable, and it is not wider than those of other 
stars, excluding the possibility that this star is a fast rotator. We 
have not a definite explanation for this anomaly, but it is possible that this is a 
very metal poor star. This possibility would be most intriguing. We notice that 
it has a large $u-y$\ colour, which might suggest a low gravity.

In addition, a few stars have very high Na abundances. Some examples are
shown in Fig.~\ref{f:figspectra}. Errors are large
for BHB stars, but anyhow star \#40227 stands out as peculiar ([Na/Fe]=1.48).
Results for RHB stars are much more robust. There are two
RHB stars that have large Na abundances and low Ba-ones. They are warmer
and brighter than typical RHB stars. They are then likely stars evolved off
the Zero Age HB. Their progenitors were likely RR Lyrae or BHB stars. There is also
one star with a very large Na excess (\#31903: [Na/Fe]=0.93). This is a very Ba-rich
star. The nature of these extremely Na-rich stars should need a special
discussion, that is deferred to a forthcoming paper.

%

\section{Conclusions}

We presented an analysis of the abundances of several elements, including
He, N, O, Na, Mg, Si, Ca, Fe, and Ba, in about a hundred stars along the
HB of NGC~1851. We observed 35 BHB stars, 1 RR Lyrae variable, 55 RHB
stars, and 13 RGB stars for comparison. Results of this analysis helped to
better understand some of the critical issues concerning the formation of
this peculiar cluster, which has a bimodal distribution of stars along the 
HB (about 2/3 RHB and 1/3 BHB), a split SGB (about 2/3 b-SGB and 1/3 f-SGB), 
and an RGB with a sequence of stars with anomalous $v-y$\ colours including
some 10\% of the stars. The stars in this anomalous RGB are Ba-rich.

The main results we obtained may be summarized as follows:
\begin{itemize}
\item RHB stars divide into two groups: the vast majority is Na-poor and O-rich;
about 10-15\% of the stars are Na-rich and moderately O-poor, most, but not all, of 
them being Ba-rich.
\item These two groups occupy distinct regions of the colour magnitude diagram,
the Na-rich stars being redder and slightly brighter than the Na-poor ones.
\item An Na-O anticorrelation exists also among BHB stars, that are on average more 
Na-rich and O-poor than RHB stars. However, there is no clear correlation with
temperature/colours.
\item The BHB stars are enriched in N, but not exceptionally so. The total CNO
abundance is unlikely to be anomalous.
\item The He abundance of the BHB is $Y=0.29\pm 0.05$. This is consistent with
both the cosmological value and a small He enhancement. This confirms the lack of
evidence for very large He enhancements within NGC~1851.
\end{itemize}

When coupled with previous knowledge about this cluster, these results clearly rule
out the explanation of the splitting of the SGB as due to variations in the total
CNO abundance. A difference in age thus remains the only plausible explanation. On the other
hand, we suggest to link the Ba-rich RHB stars with the Ba-rich RGB ones. To
explain the anomalous $v-y$\ colours, these stars should be very N-rich ([N/Fe]$\sim 1.55$).
This upper limit to the N abundance is not inconsistent with the abundance 
upper limit of [N/Fe]$<1.55$\ we obtain for one Ba-rich RHB star. 

On the whole, the observational frame suggests that most BHB stars descend
from the f-SGB stars and are old while most RHB stars descend from the b-SGB and
are young, the difference in age being of the order of 1 Gyr. However, the
correlation is possibly not one-to-one: it is in fact possible (though not at all demonstrated)
that Ba-rich RHB stars descend from f-SGB stars. If this is the case, then some of
the BHB stars and RR Lyrae variables descend from b-SGB stars, else there would be an
excess of b-SGB stars with respect to the observed RHB stars. A comparison with
the case of NGC~362, a GC with a metallicity and age very similar to that of the
young component of NGC~1851, shows that this is not unlikely. In fact, while lacking
any evident f-SGB, NGC~362 has a significant population of RR Lyrae variables and a
scatter of stars along the BHB. Remarkably, the RR Lyrae of NGC~362 and NGC~1851
are indistinguishable in the period-amplitude diagram (Szekely et al. 2007; Walker 1998), 
suggesting similar masses and luminosities. The presence of
a group of stars that might be identified with the second generation of the young component of
NGC~1851 at the cool end of the BHB would contribute explaining the lack of 
correlations between Na and O abundances with temperature along the BHB.

Several features of NGC~1851 are still unclear. The most relevant is the exact 
composition (CNO enrichment) and dating of the Ba-rich sequence. This sequence shows 
evidence for originating from polluters that experienced thermal pulses, and are then 
likely of rather small mass, but given these uncertainties its role in the formation 
scenario for this cluster is still to be understood. Progress can be obtained by
both a precise determination of the CNO content of these stars and their identification
along the SGB of NGC~1851. In addition, the match between the two RGB groups with
different Fe abundance found by Carretta et al. (2011a and 2011b) and the evolutionary
sequences requires further confirmation and clarification. Finally, the whole
scenario for the formation and evolution of this interesting cluster needs to
be put on a more sound basis.

\begin{acknowledgements}
This publication makes use of data products from
the Two Micron All Sky Survey, which is a joint project of the
University of Massachusetts and the Infrared Processing and Analysis
Center/California Institute of Technology, funded by the National
Aeronautics and Space Administration and the National Science
Foundation. This research has made use of the NASA's Astrophysical
Data System. This research has been funded by PRIN INAF
"Formation and Early Evolution of Massive Star Clusters".

We thank Sandro Villanova for having given us access to his unpublished 
results about CNO abundances in red giants of NGC~1851.

\end{acknowledgements}

\end{document}